# PASCHEN-1D: A one-dimensional fluid plasma solver with multi-mechanism surface emission and flexible external circuit coupling

Asif Iqbal[a*], Yves Heri[a], Bingqing Wang[a], Lan Jin[a], Md Arifuzzaman Faisal[a], and Peng Zhang[a+]
[a]Department of Nuclear Engineering and Radiological Sciences, University of Michigan, Ann Arbor, MI 48109, USA
Corresponding author: Asif Iqbal (asifiq@umich.edu), Peng Zhang (umpeng@umich.edu)

## Abstract

We present PASCHEN-1D (Plasma Advanced Solver with Coupled High-fidelity Emission and external Network), a one-dimensional time-dependent fluid plasma solver developed for self-consistent simulation of gas discharges and plasma breakdown with coupled electrode surface emission and flexible external circuit networks. The code solves drift-diffusion continuity equations for electrons and ions together with Poisson's equation. It is dynamically coupled to lumped RLC circuits, which self-consistently treat plasma transport, plasma-surface interaction, dielectric effects, and circuit response within a single framework. The electrode emission module includes ion-induced secondary electron emission, Fowler-Nordheim and Murphy-Good field emission, Richardson-Dushman thermionic emission, and photoemission based on a general, exact quantum mechanical emission theory. A finite-volume formulation with Kurganov-Tadmor fluxes, explicit diffusion, and fourth-order Runge-Kutta time integration is employed to ensure stable transient (sometimes ultrafast) evolution across breakdown and glow regimes. The solver is validated against multiple benchmark cases, including nanosecond pulsed dielectric-barrier discharges, DC breakdown and glow transitions, and Paschen curve construction for argon and nitrogen, with results consistent with published studies. With high-fidelity emission physics and a flexible circuit-coupling framework, PASCHEN-1D provides a versatile and efficient tool for modeling breakdown and transient discharge phenomena.

## 1. Introduction

Gas discharges and plasma breakdown play a central role in a wide range of applications [1–5], including high power electromagnetic waves and pulsed power systems [6,7], satellite communications [8,9], plasma processing [10,11], high-voltage insulation [12], radiation sources [13], particle accelerators [14,15], plasma thrusters [16–18], nuclear fusion [19,20], and emerging emission-driven plasma technologies [21]. Predictive modeling of these systems requires resolving strongly coupled physical processes, e.g., electron and ion transport, ionization kinetics, space-charge modified electric fields, surface charging, cathode and/or anode emission, external circuit interactions, etc., which often occur across disparate time scales ranging from picoseconds to microseconds. While kinetic approaches such as particle-in-cell (PIC) simulations [22] can provide high physical fidelity, their computational cost and complexity limit their routine use for parametric studies and long-time evolution of the plasmas. Fluid drift-diffusion-Poisson models [23–25] remain an essential tool for discharge modeling, particularly when augmented with physically consistent boundary conditions and transport assumptions.

A number of plasma solvers [22–28] have been developed primarily for low-temperature plasma studies. However, most publicly available implementations either neglect the external driving circuit entirely or employ highly simplified circuit representations (see Section 5 below for details). In many cases, electrode emission is treated using fixed secondary-electron boundary conditions or simplified current injection models. Consequently, self-consistent treatment of the coupling between emission mechanisms, space-charge dynamics, dielectric charging, and realistic external networks remains limited. This gap is particularly restrictive in regimes where both emission physics and circuit impedance strongly influence and determine transient plasma behavior and steady-state discharge characteristics. Another key limitation of the existing plasma solvers is that a single code often struggles to efficiently and accurately handle both ultrafast, microscopic processes (e.g. electron emission due to ultrafast laser excitation) and macroscopic plasma transport processes (e.g. evolution of a capacitively coupled plasma).

In this work, we present PASCHEN-1D (**P**lasma **A**dvanced **S**olver with **C**oupled High-fidelity **E**mission and external **N**etwork), a one-dimensional drift-diffusion-Poisson plasma solver developed to address these limitations using a unified, self-consistent framework. PASCHEN-1D combines high-resolution finite-volume transport with multiple high-fidelity electrode emission models, including ion-induced secondary emission, field emission, thermionic emission, and ultrafast photoemission, and a generalized external circuit framework capable of representing resistive, capacitive, inductive, and dielectric elements in flexible combinations. The solver allows plasma dynamics, surface charging, emission feedback, and circuit response to evolve simultaneously and consistently, which enables PASCHEN-1D to connect fluid discharge modeling with circuit-level dynamics. It provides a computationally efficient, physics-based platform for studying a wide variety of plasma systems. We present the governing equations and physical models in Section 2, numerical methods in Section 3, and validation benchmarks in Section 4. Section 5 outlines practical scope and potential avenues for improvement of this framework. Conclusion and outlook are given in Section 6.

## 2. Governing equations and physical models

PASCHEN-1D solves a self-consistent, one-dimensional drift–diffusion–Poisson system for electrons and ions in a planar discharge, coupled to an external lumped circuit and high-fidelity electrode emission models. The formulation targets plasmas where transport is well described by drift in the electrostatic field and collisional diffusion.

Throughout this section, $\mathrm{x} \in [0, \mathrm{L}]$ denotes the axial coordinate across the plasma gap, with $\mathrm{x} = 0$ at the anode and $\mathrm{x} = \mathrm{L}$ at the cathode [29].

### 2.1 Continuity equations

The electron and ion number densities, $\mathrm{n_e(x, t)}$ and $\mathrm{n_i(x, t)}$, evolve according to continuity equations of the form

$$\frac{\partial \mathrm{n_s}}{\partial \mathrm{t}} + \frac{\partial \Gamma_\mathrm{s}}{\partial \mathrm{x}} = \mathrm{S_s}, \qquad \mathrm{s} \in \{\mathrm{e}, \mathrm{i}\}, \tag{1}$$

where $\Gamma_\mathrm{s}$ is the number flux and $\mathrm{S_s}$ is the volumetric source term.

### 2.2 Drift–diffusion fluxes

The number fluxes are modeled using the standard drift–diffusion approximation,

$$\Gamma_\mathrm{i} = -\mathrm{D_i} \frac{\partial \mathrm{n_i}}{\partial \mathrm{x}} + \mu_\mathrm{i} \mathrm{n_i} \mathrm{E}, \tag{2a}$$

$$\Gamma_\mathrm{e} = -\mathrm{D_e} \frac{\partial \mathrm{n_e}}{\partial \mathrm{x}} - \mu_\mathrm{e} \mathrm{n_e} \mathrm{E}, \tag{2b}$$

where $\mathrm{E(x, t)}$ is the electrostatic field, $\mu_\mathrm{i}$ and $\mu_\mathrm{e}$ are ion and electron mobilities respectively, and $\mathrm{D_{s=e,i}}$ are diffusion coefficients. The signs reflect the opposite drift directions of electrons and ions in an electric field.

Transport coefficients may be evaluated either from user-defined analytic or empirical closures, or from local-field interpolation of externally supplied swarm-data tables, depending on the selected electron and ion kinetics models (discussed further in Section 2.4).

### 2.3 Ionization and recombination source terms

Volume source terms account for electron-impact ionization and recombination,

$$S_i = S_e = \nu_i n_e - \beta n_i n_e, \tag{3a}$$

Here $\beta$ is an effective electron-ion recombination coefficient and $\nu_i$ is the ionization frequency, defined as

$$\nu_i = \alpha(E, p)\, \mu_e |E|. \tag{4}$$

This form corresponds to a Townsend description in which ionization is driven by electron drift in the local electric field.

### 2.4 Transport and ionization coefficient models

#### 2.4.1 Electron and ion transport models

Electron transport is selected through the electron-kinetics model. In the present implementation, two electron-kinetics modes are supported: `user_defined_electron_kinetics` and `local_field_approximation`.

In the `user_defined_electron_kinetics` mode, the electron mobility $\mu_e$ and diffusion coefficient $D_e$ are evaluated from user-editable closures. This mode is intended for cases where the user wishes to prescribe analytic, empirical, or semi-empirical expressions directly, e.g., pressure-scaled fits or other problem-specific transport relations.

In the `local_field_approximation` mode, electron transport coefficients are evaluated from the local reduced electric field $(E/N)$, where $N$ is the neutral gas number density. Within this mode, the electron transport source may itself be selected using either `user_defined_equation` or `swarm_data_table_interpolation` modes. In the former case, $\mu_e$ and $D_e$ are again obtained from user-defined closures. In the latter case, the solver interpolates externally supplied swarm-data tables (e.g., from BOLSIG+ [30] or other electron-swarm solvers) [31,32] containing reduced transport coefficients, such as $(\mu_{i,e}N)_{tab}(\xi)$ and $(D_{i,e}N)_{tab}(\xi)$, where $\xi\,(x,t) = |E(x,t)|/N$. The local coefficients are then recovered through

$$\mu_{i,e}(x,t) = (\mu_{i,e}N)_{tab}/N, \tag{5a}$$
$$D_{i,e}(x,t) = (D_{i,e}N)_{tab}/N. \tag{5b}$$

This interface is solver-agnostic, i.e., any upstream Boltzmann or swarm solver capable of generating consistent transport tables as functions of $(E/N)$ may be used.

Ion transport is selected independently through `user_defined_ion_kinetics` or `local_field_ion_kinetics` modes. The positive-ion mobility and diffusion coefficient may be supplied by user-defined closures [33], by normalized ion-swarm tables [31,32], or, for diffusion, by Einstein-relation closure $D_i = \mu_i k_B T_i/e$ if only the mobility is provided by swarm data.

### 2.4.2 Electron impact-ionization models

Electron impact ionization is configured independently of the electron transport model. Two ionization pathways are currently supported: `from_townsend_alpha` and `from_ionization_frequency`.

In the `from_townsend_alpha` mode, the ionization frequency $\nu_i$ is computed from the Townsend ionization coefficient $\alpha$ and the local electron drift speed magnitude $|u_e|=\mu_e|E|$, i.e.,

$$\nu_i = \alpha|u_e| = \alpha\,\mu_e|E|. \tag{5c}$$

The Townsend coefficient $\alpha$ may be supplied in one of two ways. In the `user_defined_equation` mode, $\alpha$ is evaluated from a user-editable analytic or empirical closure. The default analytic closure is the classical Townsend form [1,2]

$$\alpha = Ap\,\exp(-Bp/|E|\,), \tag{5d}$$

where $p$ is the gas pressure in $\mathrm{Torr}$, $E$ is the local electric field magnitude, and $A$ and $B$ are gas-dependent constants. Alternative parameterizations may also be substituted directly, e.g., [1]

$$\alpha = Cp\,\exp(-D(p/|E|)^{1/2}), \tag{5e}$$

or any other suitable expression supported by the selected gas species.

In the `interpolate_from_e_over_n_table` mode, the solver interpolates tabulated reduced Townsend data $(\alpha/N)_{tab}(\xi)$, where $\xi\,(x,t) = |E(x,t)|/N$, and reconstructs the local Townsend coefficient through

$$\alpha(x,t)=(\alpha/N)_{tab}N. \tag{5f}$$

In the `from_ionization_frequency` mode, the ionization frequency $\nu_i$ is specified directly rather than through $\alpha$. As with the Townsend route, the source of $\nu_i$ may be chosen as either `user_defined_equation` or `interpolate_from_e_over_n_table`. In the table-based case, the solver interpolates tabulated reduced ionization-frequency data and reconstructs the local ionization frequency from the neutral density. This option is useful when ionization data are available directly from swarm calculations or external kinetic preprocessing, without passing through an intermediate Townsend- $\alpha$ closure.

The modular and independent structures of transport selection and ionization selection are intended to simplify extension to additional gases, alternative swarm-data sources, and more detailed kinetic closures without modification of the numerical backbone.

For table-based operation, electron tables are read from BOLSIG+-style named sections, or from equivalently structured swarm-output files, and must provide the requested reduced quantities on a consistent E/N axis. Bundled electron tables are authenticated through a manifest, and ion tables are normalized files with embedded ion-neutral identity, temperature, source, citation, and checksum provenance [31,32]. Data from other Boltzmann, swarm, or mobility databases may be used after conversion to these documented table formats.

### 2.5 Electrostatic field and Poisson equation

The electrostatic potential $\phi(x,t)$ is obtained from Poisson's equation,

$$\frac{\partial^2\phi}{\partial x^2} = -\frac{e}{\varepsilon_0}(n_i - n_e), \tag{6}$$

where $e\ (> 0)$ is the elementary charge and the charge density is defined as $\rho = e(n_i - n_e)$, with ions carrying charge $+e$ and electrons carrying charge $-e$. $\varepsilon_0$ is the permittivity of free space.

Dirichlet boundary conditions are imposed at the electrodes,

$$\phi(0,t) = V_{gap}(t), \qquad \phi(L,t) = 0, \tag{7}$$

with $V_{gap}(t)$ supplied by the external circuit model. The electric field is recovered as

$$E(x,t) = -\frac{\partial \phi}{\partial x}. \tag{8}$$

Poisson's equation is discretized using second-order finite differences on interior grid points and solved using a banded tridiagonal solver at each time step.

### 2.6 Initial and Boundary conditions

#### 2.6.1 Initial conditions

Simulations are initialized with a spatially uniform, quasi-neutral plasma,

$$n_e(x,0) = n_i(x,0) = n_0, \tag{9a}$$

and zero initial electric field,

$$\phi(x,0) = 0, \qquad E(x,0) = 0. \tag{9b}$$

The initial gap voltage is set equal to the applied voltage at $t = 0$. This initialization avoids artificial sheath formation and allows the discharge to develop self-consistently from the applied fields and emission processes.

#### 2.6.2 Boundary conditions

Boundary conditions for electrons and ions are specified independently for each electrode and each species. For a given electrode (anode or cathode) and species $s \in \{e, i\}$, the user selects one of several physically motivated boundary modes. This modular architecture allows asymmetric and emission-driven configurations to be represented without altering the interior drift–diffusion update.

The available boundary modes are:

**(a) Zero-density boundary:** In this mode, the species density is set to zero at the boundary node,

$$n_s = 0. \tag{10a}$$

This corresponds to a fully absorbing electrode with no re-emission and is appropriate for simplified absorbing or quasi-neutral configurations.

**(b) Implicit drift closure:** In this mode, the boundary density is evolved using a drift-consistent closure derived from the local drift flux [29,33]. The update is formulated implicitly to ensure stability in the presence of strong sheath electric fields.

For ions at the cathode, the boundary evolution is [29,33],

$$\frac{\partial n_i}{\partial t} = \mu_i \frac{\partial}{\partial x}\left(n_i \frac{\partial \phi}{\partial x}\right) \tag{11a}$$

while for electrons at the anode [29,33],

$$\frac{\partial n_e}{\partial t} = -\mu_e \frac{\partial}{\partial x}\left(n_e \frac{\partial \phi}{\partial x}\right). \tag{11b}$$

**(c) Electron emission boundary:** At emitting electrodes, electron emission is prescribed through a boundary number-flux condition. At the cathode, ion-induced secondary electron emission and externally driven electron emission inject electrons from the cathode into the plasma, i.e., in the $-x$ direction. The imposed cathode-side emission contribution is therefore

$$\Gamma_e(L) = -\left[\Gamma_{ext,c} + \gamma\Gamma_{i,incident}(L)\right], \tag{12a}$$

where $\Gamma_{ext,c}$ is the externally driven cathode emission flux, $\gamma$ is the ion-induced secondary electron yield, and $\Gamma_{i,incident}(L)$ is the incident ion flux at the cathode.

At the anode, electron-induced secondary electron emission is driven by electrons incident on the anode. The emitted secondary electrons are injected back into the plasma in the $+x$ direction. The anode-side electron flux is written as

$$\Gamma_e(0) = \Gamma_{ext,a} - (1-\delta)\Gamma_{e,incident}(0), \tag{12b}$$

where, $\Gamma_{ext,a}$ represents externally driven emission flux, $\delta$ is electron-induced secondary yield at the anode, and $\Gamma_{e,incident}(0)$ is the incoming component of the plasma-side electron drift-diffusion flux. Thus, when $\delta = 0$ and $\Gamma_{ext,a} = 0$, the anode behaves as a purely absorbing boundary for incoming electrons; when $\delta > 0$, a fraction of the incident electron flux is re-emitted into the plasma.

In the numerical implementation, externally driven emission may include any enabled combination of constant-current, photoemission (CW or few-cycle pulsed), field emission, and thermionic emission contributions (further discussed in Section 2.8). These contributions are summed separately for each electrode. The resulting signed emission flux, representing total electrons per time injected from the electrode into the plasma, is imposed at the electrode-plasma boundary face in the finite-volume update. The outgoing transport contribution from the plasma domain, corresponding to electrons transported from the adjacent plasma cell toward the electrode, is also retained. This allows simultaneous absorption and emission at the same electrode without introducing an artificial volumetric emission source.

### 2.7 External circuit coupling

The PASCHEN-1D framework supports self-consistent coupling between the 1D drift–diffusion–Poisson plasma model and an external lumped circuit. The coupling is bidirectional: (i) the plasma provides an instantaneous conduction and diffusion current computed from the resolved charged-particle fluxes, and (ii) the external circuit returns the instantaneous gap voltage $V_{gap}(t)$, which is imposed as a Dirichlet boundary condition on the electrostatic potential used in the Poisson solver (Section 2.5). The purpose of this module is to enable systematic studies spanning voltage-driven, current-limited, and resonant excitation conditions without modifying the plasma solver.

#### 2.7.1 Plasma representation in the circuit

The drift–diffusion model advances the ion and electron number densities and provides, at every time step, the ion and electron number-flux profiles, $\Gamma_i(x,t)$ and $\Gamma_e(x,t)$, respectively. The local conduction and diffusion current density follows the standard 1D relation

$$j_{transport}(x,t) = e(\Gamma_i(x,t) - \Gamma_e(x,t)). \tag{13}$$

To couple the plasma to an external circuit, we find the current induced in the circuit (i.e. the electrode) due to charge motion inside the plasma gap by spatially averaging the flux difference over the gap length $L$ and multiplying by the electrode area $A$:

$$I_{transport}(t) = \frac{Ae}{L}\int_0^L (\Gamma_i(x,t) - \Gamma_e(x,t))\, dx. \tag{14}$$

Eq. (14) is essentially a statement of the Ramo-Shockley theorem [34–37]. In the implementation, the integral is evaluated using the composite trapezoidal rule on the uniform mesh. This definition is robust to the choice of boundary-condition scheme because it depends on the fully resolved flux profiles rather than a face value at a single electrode.

In addition to conduction, the plasma gap behaves as a geometric capacitor with capacitance

$$C_{gap} = \frac{\varepsilon_0 A}{L}, \tag{15}$$

where $\varepsilon_0$ is the vacuum permittivity. The plasma-branch current reported by the circuit module is therefore written as the sum of conduction and displacement contributions,

$$I_{pl}(t) = I_{transport}(t) + C_{gap}\frac{dV_{gap}}{dt}, \tag{16}$$

with $V_{gap}(t)$ defined as the instantaneous voltage across the gas gap (plasma region only).

### 2.7.2 Dielectric mapping

When dielectric layers of thickness $l$ and relative permittivity $\varepsilon_r$ are present adjacent to each electrode, the circuit model follows an effective relation [33] that modifies the connection between the externally applied voltage and the plasma-gap voltage. This mapping is independent of the external circuit topology and is applied uniformly wherever a dielectric layer is present.

In this convention, the dielectric contributions are represented through two coefficients [33]

$$\alpha_d = 1 + \frac{2l}{\varepsilon_r L}, \qquad \beta_d = \frac{2el}{\varepsilon_0 \varepsilon_r L}, \tag{17}$$

which satisfy $\alpha_d \to 1$ and $\beta_d \to 0$ as $l \to 0$, ensuring a smooth reduction to the bare-electrode case.

It is convenient to define a flux integral $\Phi(t)$ such that

$$I_{transport}(t) = \frac{Ae}{L}\Phi(t) \Rightarrow \Phi(t) = \frac{L}{Ae} I_{transport}(t). \tag{18}$$

The term $\beta_d \Phi$ acts as a dielectric-mediated contribution to the effective voltage partitioning. In the "`dielectric_plasma`" configuration (no explicit external resistive, capacitive, or inductive elements), the gap-voltage dynamics can be expressed in differential form as

$$\alpha_d \frac{dV_{gap}}{dt} = \frac{dV_s}{dt} - \beta_d \Phi, \tag{19}$$

where $V_s(t)$ is the applied source voltage waveform. This relation is advanced in time and provides $V_{gap}(t)$ to the plasma solver. For $l = 0$, this reduces to the expected $dV_{gap}/dt = dV_s/dt$.

### 2.7.3 Unified external-circuit formulation

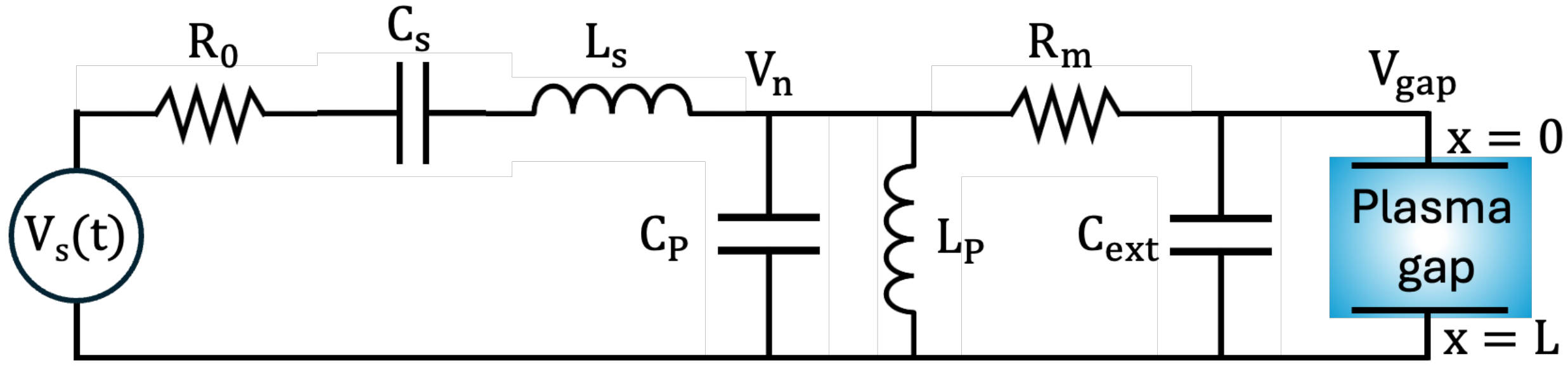

**Figure 1.** Schematic of the PASCHEN-1D external lumped-element circuit coupled to the one-dimensional plasma gap. The applied source voltage $V_s(t)$ drives the $R_0 - C_s - L_s$ series branch and feeds node $V_n$. At $V_n$, $C_p$ and $L_p$ are shunt branches, while $R_m$ connects to the plasma/electrode terminal labelled $V_{gap}$ in the figure. The load-side capacitance $C_{ext}$ and the plasma/dielectric gap branch are connected from this terminal to the return node. For bare electrodes, this terminal voltage is the internal gas-gap voltage $V_{gap}$; when dielectric layers are present, the circuit terminal voltage is related to the gas-gap voltage by the dielectric mapping. The plasma domain spans $x = 0$ to $x = L$.

Figure 1 shows the full lumped-element network used to define the plasma-circuit coupling: the $R_0 - C_s - L_s$ drive branch, the $C_p/L_p$ shunts at $V_n$, the $R_m$ load path, and the load-side capacitance $C_{ext}$ connected at the terminal labelled $V_{gap}$ in Fig. 1. The plasma is represented as a two-part element: (i) a conduction and diffusion current path provided by the drift–diffusion solver, and (ii) the geometric gap capacitance $C_{gap} = \varepsilon_0 A/L$. The circuit is driven by the prescribed source waveform $V_s(t)$. When dielectric layers are present, the circuit terminal voltage and the internal gas-gap voltage $V_{gap}$ are related by dielectric mapping; for bare electrodes the terminal voltage equals $V_{gap}$.

The maximal circuit in Fig. 1 is advanced using the state variables

$$\{V_n(t),\ V_{gap}(t),\ V_{C_s}(t),\ I_s(t),\ I_{L_p}(t)\},$$

where $V_{C_s}$ is the series-capacitor voltage, $I_s$ is the series-branch current through $R_0 - C_s - L_s$, and $I_{L_p}$ is the shunt-inductor current. The circuit dynamics follow standard KCL/KVL relations:

**Series branch (drive path):** The series inductor ($L_s$) and capacitor ($C_s$) are governed by

$$\frac{dI_s}{dt} = \frac{V_s(t) - R_0 I_s - V_{C_s} - V_n}{L_s}, \tag{20a}$$

$$\frac{dV_{C_s}}{dt} = \frac{I_s}{C_s}. \tag{20b}$$

**Shunt branch:** The shunt inductor ($L_p$) current satisfies

$$\frac{dI_{L_p}}{dt} = \frac{V_n}{L_p}. \tag{21}$$

**Node equation (Kirchhoff's Current Law at $V_n$):** The current delivered by the series branch splits into the shunt elements and the plasma branch:

$$I_s = C_p \frac{dV_n}{dt} + I_{L_p} + I_{pl}. \tag{22}$$

**Plasma-branch closure through $R_m$:** When present, $R_m$ relates the node voltage $V_n$ to the effective electrode-to-electrode voltage (including dielectric mapping). The current through $R_m$ is equated to the plasma-branch current $I_{pl}$:

$$\frac{[V_n - (\alpha_d V_{gap} + V_d)]}{R_m} = I_{transport} + C_{gap} \frac{dV_{gap}}{dt}. \tag{23}$$

Here, $V_d$ is advanced using the dielectric mapping relation $dV_d/dt = \beta_d \Phi$. For bare-metal electrodes $(l \to 0), \alpha_d \to 1,$ and $\beta_d \to 0$, recovering the usual relation $(V_n - V_{gap})/R_m = I_{transport} + C_{gap}\, dV_{gap}/dt$.

The external circuit ODEs are advanced at the plasma time step $\Delta t$; the updated $V_{gap}(t+\Delta t)$ is then imposed in the Poisson solver as the Dirichlet boundary condition at the driven electrode (Section 2.5). This arrangement keeps the plasma numerics and the lumped circuit dynamics modular while preserving a consistent current and voltage exchange at every step.

The unified maximum topology (Fig. 1) and named reduced modes (discussed in Section 2.7.5) are available with explicit-Euler and implicit-Euler circuit stepping, selected through the `circuit_time_scheme` mode. Implicit stepping is preferred when the external circuit introduces fast time scales, i.e., increased temporal stiffness, relative to the plasma time step. The unified maximum topology can additionally be advanced with a backward-Euler modified-nodal-analysis (MNA) backend (`circuit_time_scheme = mna`), where elements can be removed by assigning limiting values that make the corresponding element behave as an ideal short circuit or an ideal open circuit, as discussed in Section 2.7.4.

### 2.7.4 Reduction to simpler circuit families

The following open- or short-circuit limiting values can be used to recover reduced circuit families from the maximum topology shown in Fig. 1 when `circuit_time_scheme = mna` mode is selected.

- **No source resistor:** set $R_0 \to 0$, removing the source-side resistor as a short.
- **No shunt inductor:** set $L_p \to \infty$ (or omit $I_{L_p}$); KCL reduces to $I_s = C_p\, dV_n/dt + I_{pl}$.
- **No series inductor:** set $L_s \to 0$ (algebraic series branch) or omit the $L_s$ state.
- **No series capacitor:** set $C_s \to \infty$ (so $V_{C_s} \approx 0$) and remove the $V_{C_s}$ state, yielding a resistive drive through $R_0$.
- **No node shunt capacitor:** set $C_p \to 0$, removing the displacement current at the node.
- **No measurement/matching resistor:** set $R_m \to 0$, so $V_n \to \alpha_d V_{gap} + V_d$ (dielectric-aware, with $V_d \to 0$ in the bare metal limit $l \to 0$ when there is no surface charging).
- **No load-side capacitance:** set $C_{ext} \to 0$, removing the load-side capacitance as an open branch.
- **Plasma-only (dielectric-only) case:** remove $R_0, C_s, L_s, C_p, L_p, R_m$ and retain only the mapping between $V_s(t), \Phi(t)$, and $V_{gap}(t)$ given by Eq. (19).

Note that in the Python configuration used by PASCHEN-1D, these limits are represented by exact numerical assignments. Specifically, $R_0 = 0.0$, $L_s = 0.0$, and $R_m = 0.0$ represent short-circuit limits; $C_p = 0.0$ and $C_{ext} = 0.0$ remove the corresponding shunt capacitances as open branches; and $L_p = \mathrm{np.inf}$ and $C_s = \mathrm{np.inf}$ represent the shunt-inductor open-circuit limit and the series-capacitor short-circuit limit, respectively.

### 2.7.5 Implemented circuit modes (configuration layer)

While the external-circuit formulation is presented above in a unified manner using the maximal network shown in Fig. 1, the implementation also provides a set of circuit modes that correspond to commonly used experimental and modeling configurations. Each of these modes represents a well-defined reduction of the maximal circuit obtained by removing selected elements. They are included to simplify configuration and improve readability.

In the current PASCHEN-1D implementation, the available circuit topologies are:

- `dielectric_plasma`: dielectric-plasma coupling only, with dielectric mapping and no explicit lumped external $R, C,$ or $L$ branch elements.
- `R0_Cp`: voltage source feeding the discharge through a series resistor $R_0$, with a shunt capacitor $C_p$ at the plasma-side node.
- `R0_Cp_Rm`: the `R0_Cp` topology with an additional series plasma-branch resistor $R_m$.
- `R0_Rm_Cext`: resistive drive through $R_0$ and $R_m$ with an optional load-side stray capacitance $C_{ext}$ at the gap/load terminal.
- `R0_Cs_Cp`: the `R0_Cp` topology with an added series capacitor $C_s$ in the drive branch, representing capacitive source coupling.
- `R0_Cs_Cp_Rm`: the `R0_Cs_Cp` topology with an additional plasma-branch resistor $R_m$.
- `R0_Cs_Ls_Cp`: the `R0_Cs_Cp` topology with an added series inductor $L_s$ in the drive branch, enabling inductive or resonant drive behavior.
- `R0_Cs_Ls_Cp_Rm`: the `R0_Cs_Ls_Cp` topology with plasma-branch resistor $R_m$.
- `R0_Cs_Ls_Cp_Lp`: the `R0_Cs_Ls_Cp` topology with an added shunt inductor $L_p$ at the plasma-side node, representing matching or parasitic inductive paths.
- `R0_Cs_Ls_Cp_Lp_Rm_Cext`: maximum topology shown in Fig. 1.

All circuit modes use the same plasma-circuit current and voltage exchange interface. A systematic verification of the implemented circuit coupling, including the independent response of each lumped element and its impact on the plasma gap voltage and discharge current, is provided in Appendix A.

### 2.8 Surface emission models

To enable simulations in regimes where charged-particle injection from surfaces is non-negligible (e.g., photo-triggered breakdown, field-assisted initiation, or sustained discharge operation), the present solver includes a high-fidelity electrode-emission module. The emission module returns the emission current density $J_{emit}(t, V_{gap}, E_{electrode}, \Delta t)$. In the drift–diffusion update, this is applied through the electron number-flux boundary condition in Eq. (12) by identifying the externally driven emission number flux as

$$\Gamma_{ext}(t) = \frac{J_{emit}(t, V_{gap}, E_{electrode}, \Delta t)}{e}. \tag{24}$$

The code supports the following emission modes, selected by the user through the simulation configuration:

a) **Ion induced secondary emission (cathode):** Ion-induced secondary emission at the cathode is modeled through a yield coefficient $\gamma$, such that the emitted electron number flux is proportional to the ion flux incident on the electrode, $\Gamma_{\mathrm{secondary}} = \gamma\Gamma_{\mathrm{i}}$. This contribution is incorporated self-consistently in the electron-emission boundary condition and may be combined with externally driven emission mechanisms.
b) **Electron induced secondary emission (anode):** Electron-induced secondary emission from the anode is modeled through a yield coefficient $\delta$, with two available options:
   - Constant yield model: $\delta$ is a user-defined constant.
   - Vaughan's model [38]: $\delta = \delta(E_p)$ according to Vaughan's empirical model of electron induced secondary electron yield [38], where the approximate average electron impact energy $E_p$ is computed as

$$E_p = \frac{m_e u_{inc}^2}{2e} + 2T_e(eV), \qquad u_{inc} = \mu_e E(0). \tag{25}$$

   Here, the first term represents the $(-x)$ directed (normal) drift kinetic energy of incoming electrons. The second term represents the half-range, flux-weighted mean thermal kinetic energy of a Maxwellian electron distribution incident on a surface, which equals $2T_e(eV)$ rather than the bulk average value $3/2T_e(eV)$(see Eq. 2.4.11 in [2]).

c) **Prescribed (constant) emission**: A user-defined constant emission current density is applied over a specified time window. This option is primarily intended for controlled numerical experiments (e.g., assessing breakdown sensitivity to seed charge) and for representing simplified source injection when detailed surface physics is not required.
d) **Field emission:** Field emission is modeled as a function of the instantaneous electric field at the electrode surface. Two field-emission models are supported. The Fowler–Nordheim (FN) model [39,40] provides a simplified description based on electron tunneling through a triangular surface barrier and is computationally efficient for parametric and qualitative studies. In addition, the Murphy–Good (MG) formulation [41] is available, which incorporates image-charge (Schottky-Nordheim) barrier lowering and a more accurate description of tunneling through the rounded surface potential. These models are appropriate when electrode-emission is dominated by strong local electric fields.
e) **Thermionic emission**: Thermionic emission is included through the temperature- and work-function-based Richardson–Dushman model [42,43] using a prescribed electrode temperature. The present implementation is time-independent in temperature (i.e., the electrode temperature is treated as a fixed parameter) and provides a thermionic emission contribution for cases where thermionic emission is expected to contribute to discharge initiation or maintenance.

f) **Pulsed photoemission**: For photoemission-driven studies [21], a quantum photoemission model [44–48] is supported. The emission current density pulse $J_{emit}(t)$ is evaluated on a fine, user-defined picosecond time grid around the laser arrival time and then embedded into the discharge simulation as an efficient runtime emitter. During the plasma time advance, the precomputed pulse is sampled (or time-averaged over $\Delta t$) and applied as the electrode boundary injection. This approach amortizes the cost of the quantum mechanical calculation while preserving accurate short-time injection dynamics relative to the plasma time step.

The code supports the activation of multiple emission models simultaneously. In that case, the total externally driven emission current density is computed as the sum of the contributions from each active mechanism, i.e., $J_{emit} = \sum_k J_{emit}^{(k)}$.

## 3. Numerical methods

PASCHEN-1D employs a finite-volume formulation on a uniform one-dimensional grid, combined with an explicit high-resolution scheme for transport and an implicit Poisson solver for the electrostatic potential. The plasma species are advanced explicitly over a user-selected time step; when adaptive substepping is enabled, that user-selected time step is internally divided into smaller transport substeps selected from the local drift-CFL estimate [49–51]. Circuit variables may be updated using explicit Euler, implicit Euler, or the modified-nodal-analysis (MNA) backend, as discussed in Section 2.7.3.

The transport hot loops (i.e., the most computationally intensive transport loops in Python) can be executed with either the default `NumPy` backend or the optional `Numba` backend [52]. The `Numba` backend preserves the same finite-volume update, but reduces runtime by just-in-time (JIT) compiling [52] the corresponding arithmetic operations to machine code before execution.

### 3.1 Spatial discretization

The plasma domain $x \in [0, L]$ is discretized using a uniform grid with $N_x$ points,

$$x_j = j\,\Delta x, \qquad \Delta x = \frac{L}{N_x - 1}, \qquad j = 0, \dots, N_x - 1. \tag{26}$$

Spatial derivatives are approximated using second-order finite differences in the interior and one-sided differences at the boundaries where required.

### 3.2 Finite-volume form of the continuity equations

The continuity equations are written in conservative form as in Eq. (1). In discrete form, the update for each species at grid point j reads

$$\frac{dn_{s,j}}{dt} = -\frac{\Gamma_{s,j+1/2} - \Gamma_{s,j-1/2}}{\Delta x} + S_{s,j}, \tag{27}$$

where $\Gamma_{s,j\pm1/2}$ are numerical fluxes evaluated at cell faces.

### 3.3 Drift flux discretization: Kurganov–Tadmor scheme

The drift component of the flux is treated using a Kurganov–Tadmor (KT) central-upwind scheme [53], chosen for its robustness in strongly advective regimes without requiring Riemann solvers.

For ions, the drift flux function is

$$f_i(n_i, E) = \mu_i n_i E, \tag{28a}$$

and for electrons,

$$f_e(n_e, E) = -\mu_e n_e E. \tag{28b}$$

At each cell face j+1/2, the KT numerical flux is constructed as

$$\Gamma_{j+1/2}^{drift} = \frac{1}{2}[f(n_L, E_{j+1/2}) + f(n_R, E_{j+1/2})] - \frac{1}{2}a_{j+1/2}(n_R - n_L), \tag{29}$$

where $n_L$ and $n_R$ are reconstructed left and right states, and

$$a_{j+1/2} = \max\left|\frac{\partial f}{\partial n}\right| = |\mu_s E_{j+1/2}| \tag{30}$$

is the local characteristic speed.

In the present implementation, first-order reconstruction is used (piecewise constant states), which prioritizes numerical robustness over formal spatial accuracy. This choice is deliberate, as breakdown and emission-driven transients can produce extremely steep gradients.

### 3.4 Explicit diffusion term

Diffusion is treated explicitly and separately from the drift flux. The diffusion contribution to the flux divergence is written as

$$\frac{\partial n_s}{\partial t}|_{diff} = \frac{D_s}{\Delta x^2}\left(n_{s,j+1} - 2n_{s,j} + n_{s,j-1}\right), \tag{31}$$

which corresponds to a second-order central difference approximation of $\frac{\partial^2 n_s}{\partial x^2}$ .

This explicit treatment is stable provided the time step satisfies a diffusion-related constraint, which is monitored through CFL diagnostic [49–51] (see Section 3.7).

### 3.5 Time integration: fourth-order Runge–Kutta

Time integration of the semi-discrete system is performed using a classical fourth-order Runge–Kutta (RK4) method. For each species,

$$n_s^{n+1} = n_s^n + \frac{\Delta t}{6}(k_1 + 2k_2 + 2k_3 + k_4), \tag{32}$$

where the stage derivatives $k_m$ are evaluated from the right-hand side of Eq. (27), including drift, diffusion, and source terms.

All source terms (ionization, recombination) are evaluated explicitly using the plasma state at the corresponding RK stage.

### 3.6 Picard iteration for density–Poisson coupling

Because the density boundary conditions depend explicitly on the electric field, the Poisson equation and the density updates are weakly nonlinear and coupled. PASCHEN-1D resolves this coupling using a fixed-point (Picard) iteration at each time step:

1. Electron and ion densities are first advanced explicitly using the Kurganov–Tadmor flux scheme with fourth-order Runge–Kutta time integration.
2. Boundary conditions are applied using the electric field from the previous iteration.
3. The Poisson equation is solved to obtain an updated potential and field.
4. Steps (2)–(3) are repeated until convergence in $\phi$ is achieved.

This procedure ensures consistency between density boundary conditions and the electrostatic field without introducing a fully implicit nonlinear solver.

### 3.7 Stability monitoring

Although the time step $\Delta t$ is user specified, PASCHEN-1D monitors a CFL-like diagnostic [49–51] based on the maximum drift speed,

$$\mathrm{CFL} = \max_x \left( \frac{|\mu_e E|\Delta t}{\Delta x}, \frac{|\mu_i E|\Delta t}{\Delta x} \right). \tag{33}$$

This quantity is recorded throughout the simulation as a stability and diagnostic metric. In fixed-step operation, it is used only as a warning and post-run diagnostic. When adaptive substepping is active, this stability condition is enforced at the substep level: the solver estimates the required number of substeps for each $\Delta t$, records the selected substep count and effective substep size, and accepts or rejects the step according to the user-selected overflow policy.

## 4. Model validation and benchmark cases

To establish the accuracy and physical fidelity of the PASCHEN-1D framework, we validate the solver in this section against well-documented benchmark problems, e.g., nanosecond pulsed discharges, dielectric effects, DC Townsend and glow discharge, breakdown, and photoemission driven strongly transient plasma dynamics. All benchmark cases are reproduced using parameters reported in the original references, without tuning or empirical fitting beyond those explicitly stated.

### 4.1 Nanosecond pulsed dielectric-barrier discharge

As a first validation test, we reproduce the nanosecond pulsed dielectric-barrier discharge reported by Adamovich et al. [33]. This case provides a benchmark for transient plasma kinetics, dielectric charging, and self-consistent field–particle coupling under nanosecond-scale excitation. All geometric parameters, dielectric properties, and excitation conditions are taken from [33], including the planar discharge geometry, quartz dielectric layers ($\varepsilon_r$=4.3), and a Gaussian voltage pulse with peak amplitude $V_{peak}$=20 kV and duration $\tau$=15ns.

Figure 2 compares PASCHEN-1D results with the reference trends. The applied voltage waveform is reproduced (Fig. 2(a)). The temporal evolution of the mid-gap electron density exhibits a rapid avalanche onset and saturation timing (Fig. 2(b)) consistent with that reported by Adamovich et al. [33]. Spatial profiles of electron density and electric field at selected times (Figs. 2(c,d)) demonstrate the formation of a dense plasma region, strong space-charge–driven field screening, followed by a dielectric-mediated field redistribution and quasi-steady state.

The agreement in breakdown timing, density magnitude, and spatial structure to those reported in [33] supports that PASCHEN-1D accurately captures the coupled drift–diffusion–Poisson dynamics and dielectric charging effects governing nanosecond pulsed discharges.

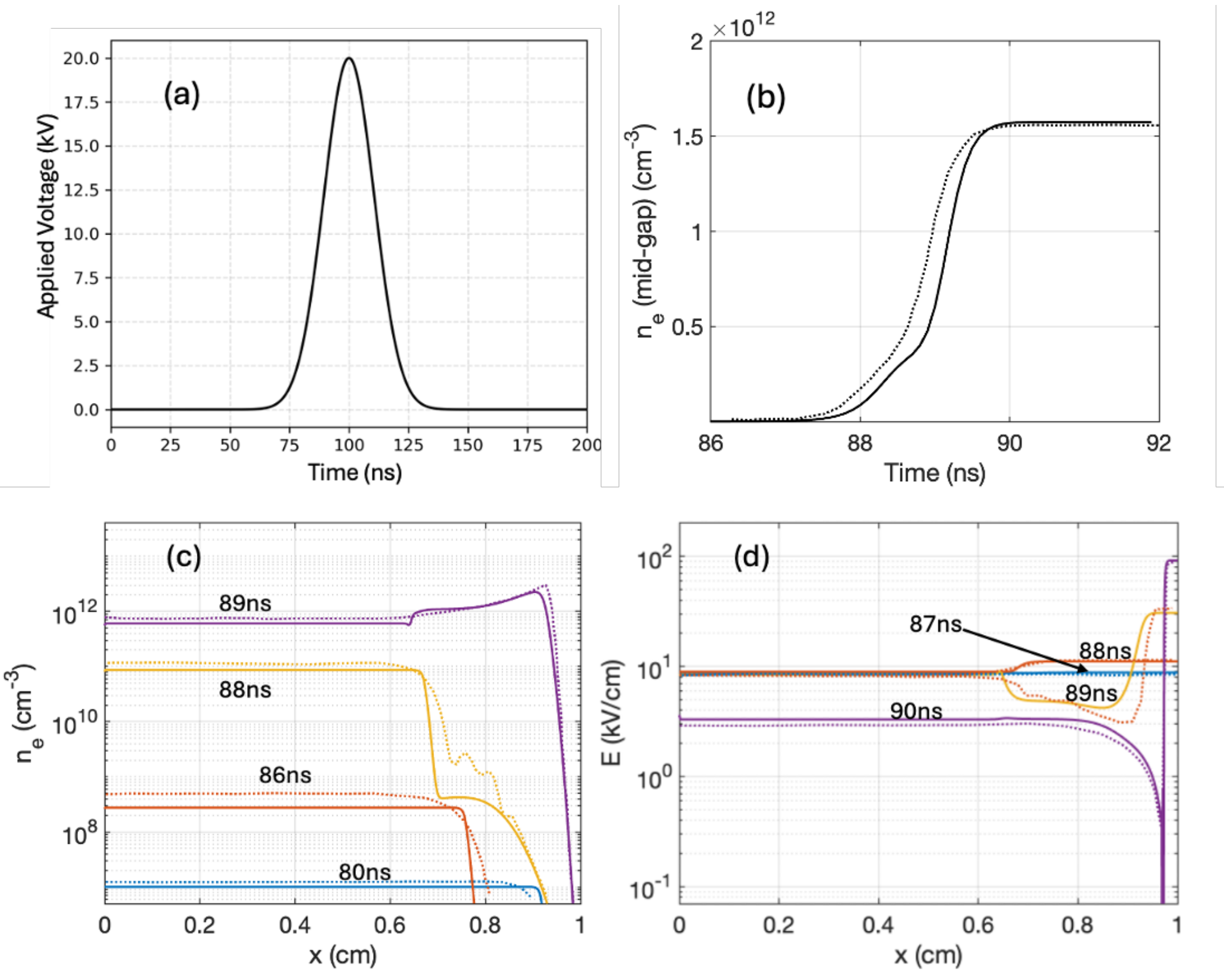

**Figure 2.** Reproduction of nanosecond pulsed dielectric-barrier discharge results from Adamovich *et al.* [33] using the PASCHEN-1D solver. (a) Applied Gaussian voltage pulse $V_s(t)$ with peak amplitude $V_{peak} = 20\,kV$ and pulse width $\tau = 15\,ns$. (b) Temporal evolution of the electron density at the mid-gap location $x = L/2$ (solid curve), showing close agreement with Fig. 8(b) of Ref. [33] (dashed curve) in avalanche onset and subsequent saturation behavior. (c) Spatial profiles of the electron number density $n_e(x)$ at selected times (solid curves), corresponding to Fig. 7(a) of Ref. [33] (dashed curves), showing the transition from localized avalanche growth to a quasi-neutral plasma. (d) Corresponding spatial electric-field profiles $E(x)$ (solid curves), reproducing Fig. 7(b) of Ref. [33] (dashed curves) and illustrating strong field screening due to space-charge accumulation and dielectric charging. All geometric, dielectric, and operating parameters (gap length $L = 1cm$, electrode area $A = 10\,cm^2$, dielectric thickness $l = 1.75mm$, permittivity $\epsilon_r = 4.3$, and nitrogen gas pressure $p = 60$ Torr) match those reported

in the reference. We use $\Delta x = 10^{-5}$m, $\Delta t = 10^{-12}$s, and a slope limiter $\theta = 1.01$ in the Kurganov-Tadmor Scheme [53]. The close agreement in breakdown timing, density magnitude, and field evolution confirms the fidelity of the coupled drift–diffusion–Poisson and dielectric-aware circuit formulation implemented in PASCHEN-1D.

### 4.2 Numerical convergence and resolution verification

Here, we present a numerical convergence study of PASCHEN-1D for a DC nitrogen discharge at $\mathrm{pd} = 1$ Torr cm. The convergence study focuses on a physically meaningful observable, i.e., the quasi-steady electron density at the mid-gap location, as the spatial and temporal discretization are refined.

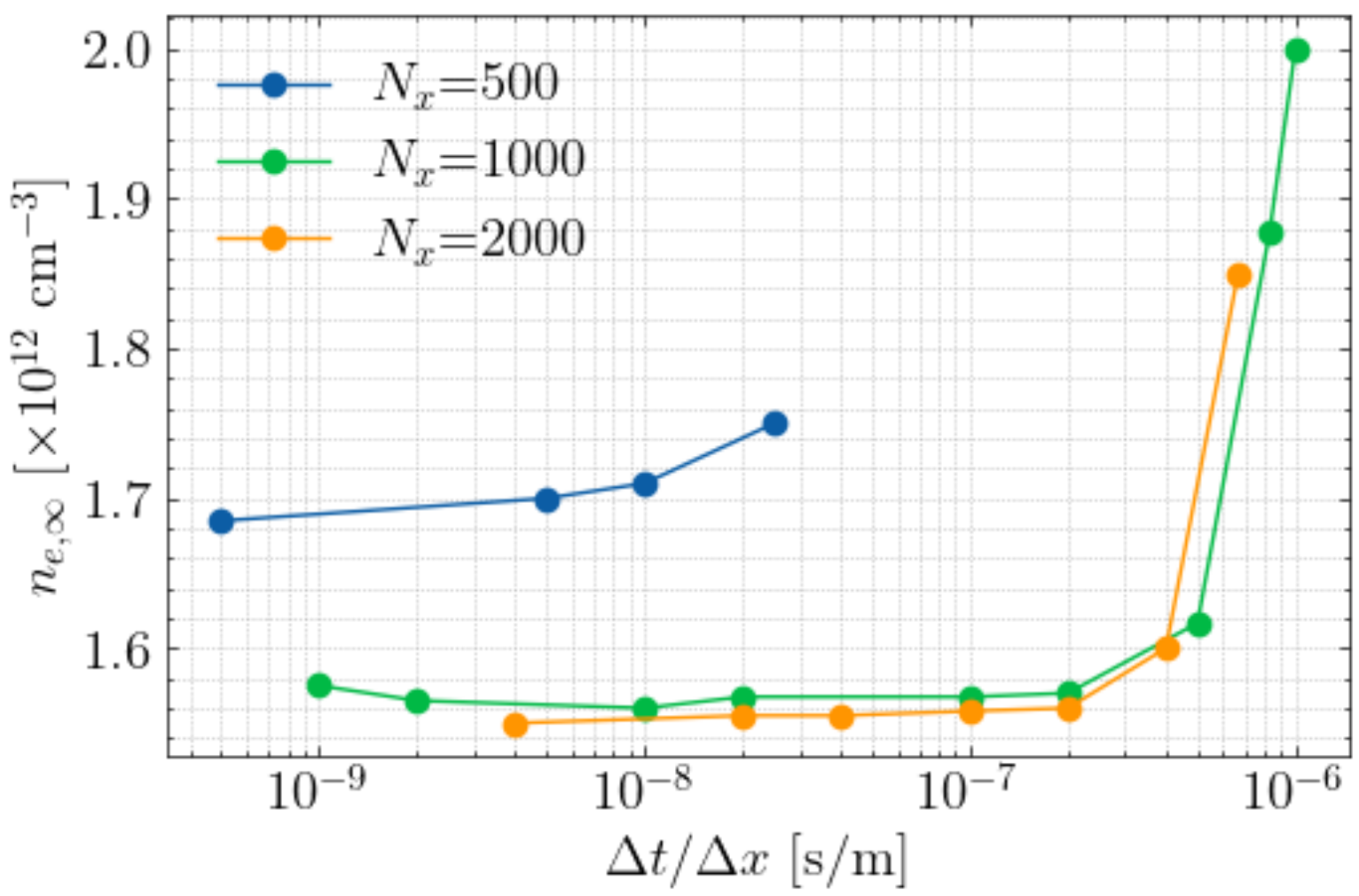

**Figure 3.** Numerical convergence study for a DC nitrogen discharge at $\mathrm{pd} = 1$ Torr cm. Electron number density at the mid-gap location ($x = L/2$) evaluated at late time as a function of the discretization ratio $\Delta t/\Delta x$, shown for three spatial resolutions ($N_x = 500, 1000, 2000$). Convergence toward a resolution-independent plateau is observed for sufficiently small $\Delta t/\Delta x$, confirming numerical consistency of the coupled drift–diffusion–Poisson and dielectric-aware circuit formulation used in PASCHEN-1D.

Figure 3 shows the electron number density at $x = L/2$ after the discharge has reached a quasi-steady state, plotted as a function of the ratio $\Delta t/\Delta x$ for three spatial resolutions ($N_x = 500, 1000, 2000$). For a fixed value of $N_x$, sufficiently small values of $\Delta t/\Delta x$, the solution converges toward a resolution-independent plateau, consistent with the coupled drift–diffusion update, Poisson solver, and dielectric-aware boundary treatment. Coarser spatial discretization (e.g., $N_x = 500, 1000$) exhibits mild overprediction of the electron density relative to $N_x = 2000$, consistent with increased numerical diffusion and reduced resolution of the sharp ionization front during breakdown. For larger values of $\Delta t/\Delta x$ , all spatial resolutions show a rapid increase in the predicted electron density, indicating insufficient temporal resolution of the fast ionization dynamics.

All benchmark results presented in this study are obtained using discretization within the converged regime, so that the reported agreement with reference results reflects physical modeling rather than numerical artifacts.

### 4.3 DC breakdown and glow discharge transition

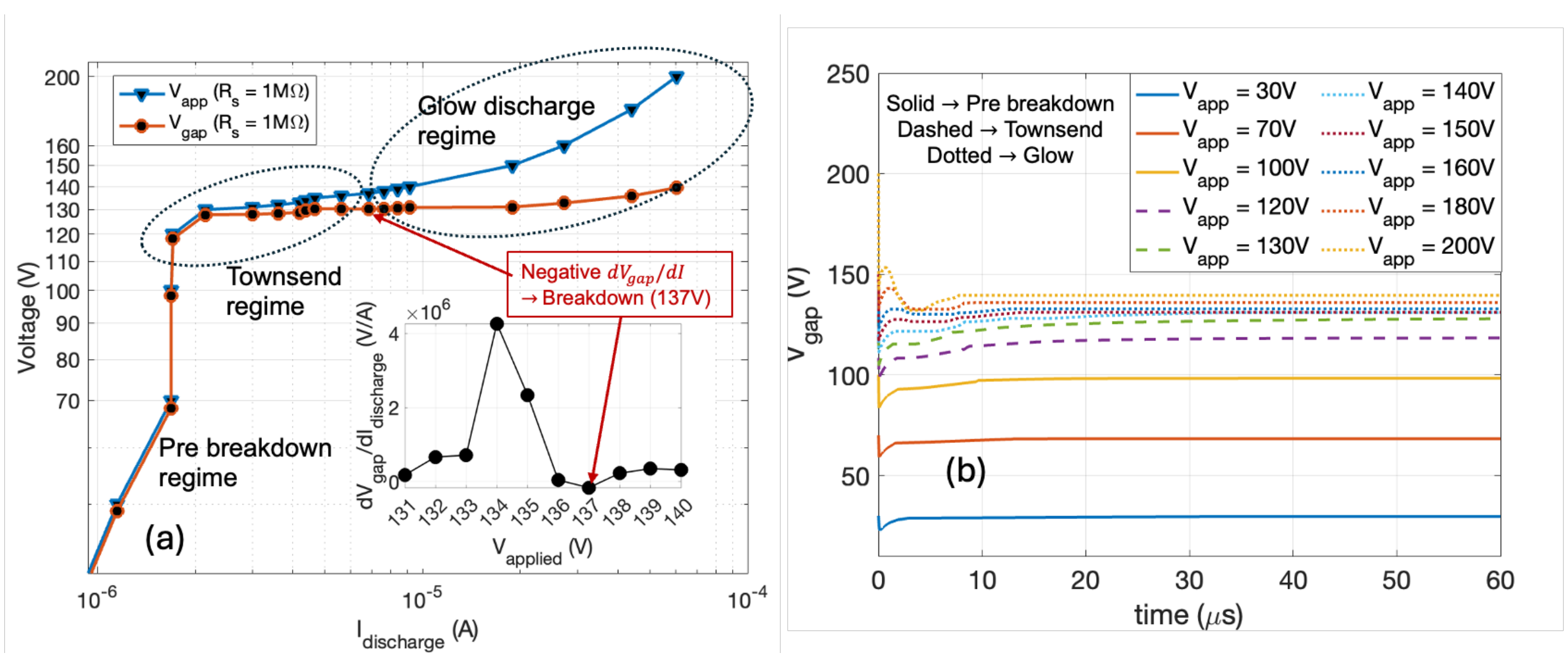

**Figure 4.** DC discharge characteristics in argon at $\mathrm{pd} = 1\ \mathrm{Torr\ cm}$ obtained using PASCHEN-1D with a series ballast resistor $R_0 = 1\ \mathrm{M\Omega}$ and no dielectric layer. **(a)** Applied voltage $V_{app}$ (blue) and steady-state gap voltage $V_{gap}$ (orange) plotted against the discharge current, revealing distinct pre-breakdown, Townsend, and glow discharge regimes. The onset of breakdown is identified by the emergence of a negative differential slope $dV_{gap}/dI < 0$, occurring at $V_{app} \approx 137\ \mathrm{V}$. The inset highlights the derivative $dV_{gap}/dI$, confirming the breakdown criterion. **(b)** Temporal evolution of the gap voltage $V_{gap}(t)$ for selected applied voltages spanning the three discharge regimes. In all cases, the system relaxes to a quasi-steady-state plateau on microsecond timescales. The steady-state values extracted from these plateaus are used to construct the current–voltage characteristics in panel (a). For argon we used transport coefficients [1]: $\mu_e = 29.3/p, \mu_i = 0.15/p$, $D_{i,e} = \mu_{i,e} k_B T_{i,e}/e$ ($k_B$ = Boltzmann coefficient) and rate coefficients [2]: $\beta = 2 \times 10^{-13}, A = 11.5\ \mathrm{cm^{-1}\ Torr^{-1}}$ and $B = 176\ \mathrm{V\ cm^{-1}\ Torr^{-1}}$.

This benchmark case assesses the ability of PASCHEN-1D to reproduce the classical DC breakdown under a constant $V_{app}$ and glow discharge behavior in a gas discharge with external current limiting. Simulations are performed for argon at a pressure–gap product of $\mathrm{pd} = 1\ \mathrm{Torr\ cm}$, using bare metal electrodes and a series ballast resistor $R_s = 1\ \mathrm{M\Omega}$. No dielectric layer or additional circuit elements are included, so that the discharge dynamics are governed by the coupled drift–diffusion–Poisson plasma model and the external resistive load.

Figure 4(a) shows the steady-state current–voltage (I-V) characteristics obtained by sweeping the applied voltage $\mathrm{V_{app}}$ and recording the corresponding gap voltage $\mathrm{V_{gap}}$ and discharge current $\mathrm{I_{discharge}}$ after convergence. At low applied voltages, the discharge operates in the pre-breakdown regime, where the plasma density remains negligible and $\mathrm{V_{gap}} \approx \mathrm{V_{app}}$. As the applied voltage is increased, the system transitions into the Townsend regime, characterized by weak ionization, finite conduction current, and a gradual reduction of $\mathrm{V_{gap}}$ due to space-charge effects.

A breakdown occurs at $\mathrm{V_{app}} \approx 137\ \mathrm{V}$, marked by the appearance of a negative differential slope $\mathrm{dV_{gap}/dI} < 0$. This behavior, highlighted in the inset of Fig. 4(a), is a standard signature of DC breakdown [1] and reflects the rapid increase in plasma conductivity associated with avalanche ionization. Beyond this point, the discharge transitions into the glow discharge regime, where the gap voltage becomes weakly dependent on the applied voltage and is instead regulated by plasma processes and the external ballast resistor. The applied voltage continues to rise with increasing current, while the gap voltage remains comparatively clamped, consistent with classical glow discharge behavior.

The temporal evolution of the gap voltage for representative applied voltages is shown in Fig. 4(b). For each applied voltage, $\mathrm{V_{gap}(t)}$ exhibits a short transient followed by relaxation to a stable plateau on microsecond timescales. Therefore, the discharge reaches a quasi-steady-state equilibrium for all applied voltages considered. The current–voltage data in Fig. 4(a) are extracted exclusively from these steady-state plateaus, ensuring that the observed negative differential resistance and regime transitions are physical and not artifacts of transient dynamics or numerical instability.

It is important to note that the simulated I–V characteristics do not exhibit a pronounced normal glow plateau, which can be attributed to the one-dimensional formulation of the model. In experiments, the normal glow regime arises from lateral expansion of the active cathode area [1], allowing the discharge current to increase at nearly constant cathode fall voltage. This mechanism cannot be captured in a 1D model, where the cathode is assumed to be uniformly active by construction and plasma column expansion is excluded [54].

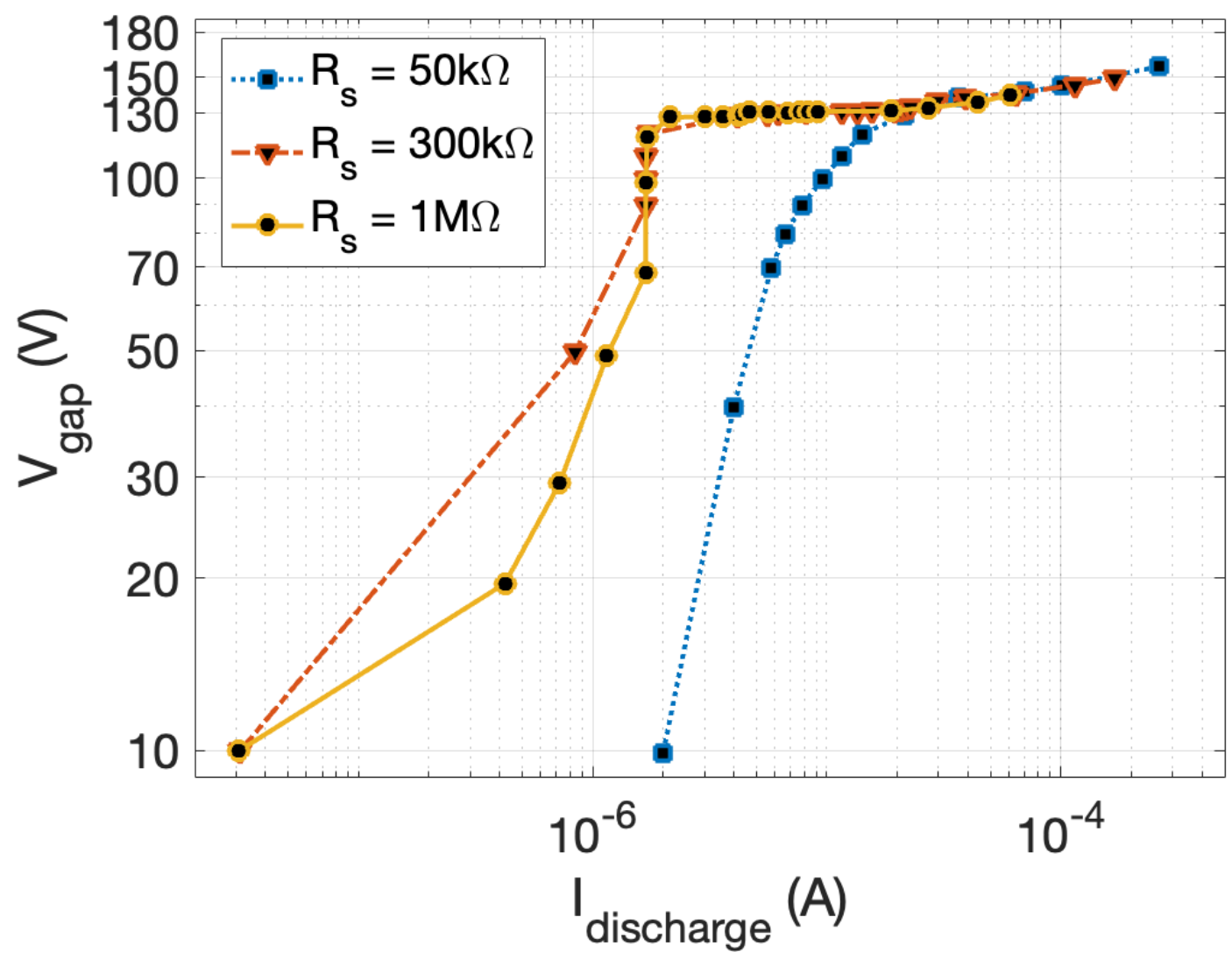

**Figure 5.** Effect of ballast resistance on the post-breakdown DC discharge characteristics in argon at $\mathrm{pd} = 1\ \mathrm{Torr\ cm}$. Shown are the gap voltage $V_{gap}$ as a function of discharge current for three series resistances: $R_s = 50\ \mathrm{k\Omega}, 300\ \mathrm{k\Omega}$, and $1\ \mathrm{M\Omega}$. For moderate ballast resistance ($R_s = 50\ \mathrm{k\Omega}$), the discharge transitions directly from breakdown into the abnormal glow regime, with no sustained normal-glow voltage plateau. Increasing the ballast resistance progressively flattens the post-breakdown I–V characteristic and introduces a weak negative differential slope (as observed in Fig. 4(a)); however, this behavior arises from current limitation by the external circuit rather than from cathode-area expansion, which is not captured in the one-dimensional model. For argon we used transport coefficients [1]: $\mu_e = 29.3/p, \mu_i = 0.15/p, D_{i,e} = \mu_{i,e} k_B T_{i,e}/e$ and rate coefficients [2]: $\beta = 2 \times 10^{-13}, A = 11.5\ \mathrm{cm^{-1}\ Torr^{-1}}$ and $B = 176\ \mathrm{V\ cm^{-1}\ Torr^{-1}}$.

In a 1D model, post-breakdown behavior is strongly influenced by the external ballast resistance, as shown in Figure 5. At moderate ballast values (e.g., $R_s \sim 50\ \mathrm{k\Omega}$), the discharge transitions directly from breakdown into the abnormal glow regime, bypassing the normal glow plateau. For sufficiently large ballast resistances (e.g., $R_s \sim 1\ \mathrm{M\Omega}$), a weak voltage plateau and negative differential slope can be recovered; however, this behavior is circuit-limited rather than due to true cathode area expansion [54].

### 4.4 Paschen curve and the role of ion-induced secondary electron emission

Figure 6 presents the Paschen curves for argon and nitrogen obtained using the PASCHEN-1D solver, alongside the corresponding experimental data reported in [2]. Breakdown voltages were extracted following the procedure described in Section 4.3 (i.e., by identifying the onset of a negative differential gap-voltage response with respect to discharge current). Over the range of reduced gap distances considered, PASCHEN-1D reproduces both the minimum breakdown voltage and the overall shape of the experimental Paschen curves for both gases with good agreement.

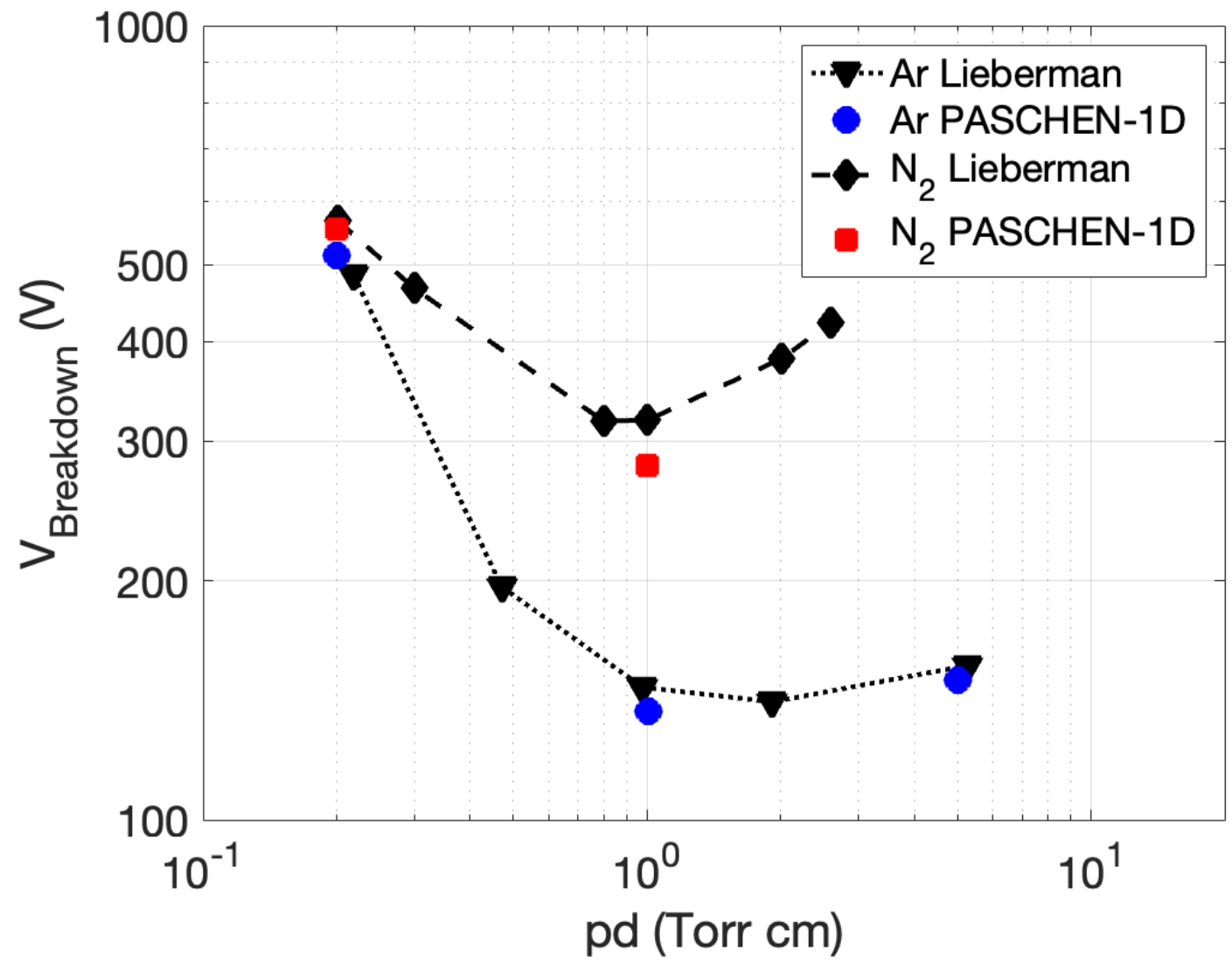

**Figure 6.** Paschen curves for argon and nitrogen obtained using the PASCHEN-1D solver for $d = L = 1\text{cm}$ and $0.24 < p\,(\text{Torr}) < 5$, compared against experimental data reported in [2]. The breakdown voltage $V_{Breakdown}$ is shown as a function of the pressure–gap product $pd$. The close agreement near the Paschen minimum and across both branches demonstrates the ability of PASCHEN-1D to accurately capture gas-dependent breakdown behavior under DC conditions using consistent transport and emission models. For argon we used transport coefficients [1]: $\mu_e = 29.3/p, \mu_i = 0.15/p$, $D_{i,e} = \mu_{i,e} k_B T_{i,e}/e$ and rate coefficients [2]: $\beta = 2 \times 10^{-13}, A = 11.5\ \text{cm}^{-1}\ \text{Torr}^{-1}$ and $B = 176\ \text{V cm}^{-1}\ \text{Torr}^{-1}$. For nitrogen we used transport coefficients [33]: $\mu_e = 30.4/p, \mu_i = 0.209/p$, $D_{i,e} = \mu_{i,e} k_B T_{i,e}/e$ and rate coefficients [2]: $\beta = 2 \times 10^{-13}, A = 11.8\ \text{cm}^{-1}\ \text{Torr}^{-1}$ and $B = 325\ \text{V cm}^{-1}\ \text{Torr}^{-1}$.

While this agreement shows the capability of the solver to recover classical breakdown trends, a closer examination reveals that the match cannot be obtained using a single, constant value of the ion-induced secondary electron emission yield, $\gamma$, across the entire $pd$ range. This limitation becomes particularly apparent on the low-$pd$ (left) branch of the Paschen curve.

To illustrate this point, Fig. 7(a) compares the experimental argon Paschen curve from [2] with predictions from the classical Paschen's law [1,2,4]

$$V_{Breakdown} = \frac{Bpd}{\ln(Apd) - \ln\left(\ln\left(1 + \frac{1}{\gamma}\right)\right)} \tag{34}$$

using fixed Townsend ionization coefficient parameters $\mathrm{A} = 11.5\,\mathrm{cm}^{-1}\mathrm{Torr}^{-1}$ and $\mathrm{B} = 176\,\mathrm{V\,cm}^{-1}\mathrm{Torr}^{-1}$, but varying $\gamma$. The results show that the left branch is extremely sensitive to the assumed value of $\gamma$, with small changes producing order-of-magnitude variations in the predicted breakdown voltage. At low $\mathrm{pd}$, the mean-free path between ionizations, $\lambda_{\mathrm{ion}} = 1/\alpha$ often becomes comparable to the electrode separation, resulting in very few ionization collisions across the gap. For instance, at $\mathrm{p} = 0.24\,\mathrm{Torr}$ and $\mathrm{d} = 1\mathrm{cm}$ (i.e., $\mathrm{pd} = 0.24\,\mathrm{Torr\,cm}$, corresponding to the leftmost points in the Paschen curves in Fig. 7(a)), $\lambda_{\mathrm{ion}}$ evaluated at the simulated breakdown voltages for $\gamma = 0.08, 0.07,$ and $0.06$ ($\mathrm{V_{Breakdown}} \approx 415\mathrm{V}, 775\mathrm{V},$ and $14.5\mathrm{kV},$ respectively) are approximately $4.01\mathrm{mm}, 3.83\mathrm{mm},$ and $3.63\mathrm{mm}$, respectively. This corresponds to only $2.49, 2.61,$ and $2.75$ ionization events across the gap, respectively. Therefore, at low $\mathrm{pd}$, breakdown becomes increasingly sensitive to surface emission processes rather than volumetric ionization (the $\alpha$-process) alone. In contrast, the high-$\mathrm{pd}$ (right) branch of the Paschen curve is dominated by the $\alpha$-process, i.e., parameters $\mathrm{A}$ and $\mathrm{B}$.

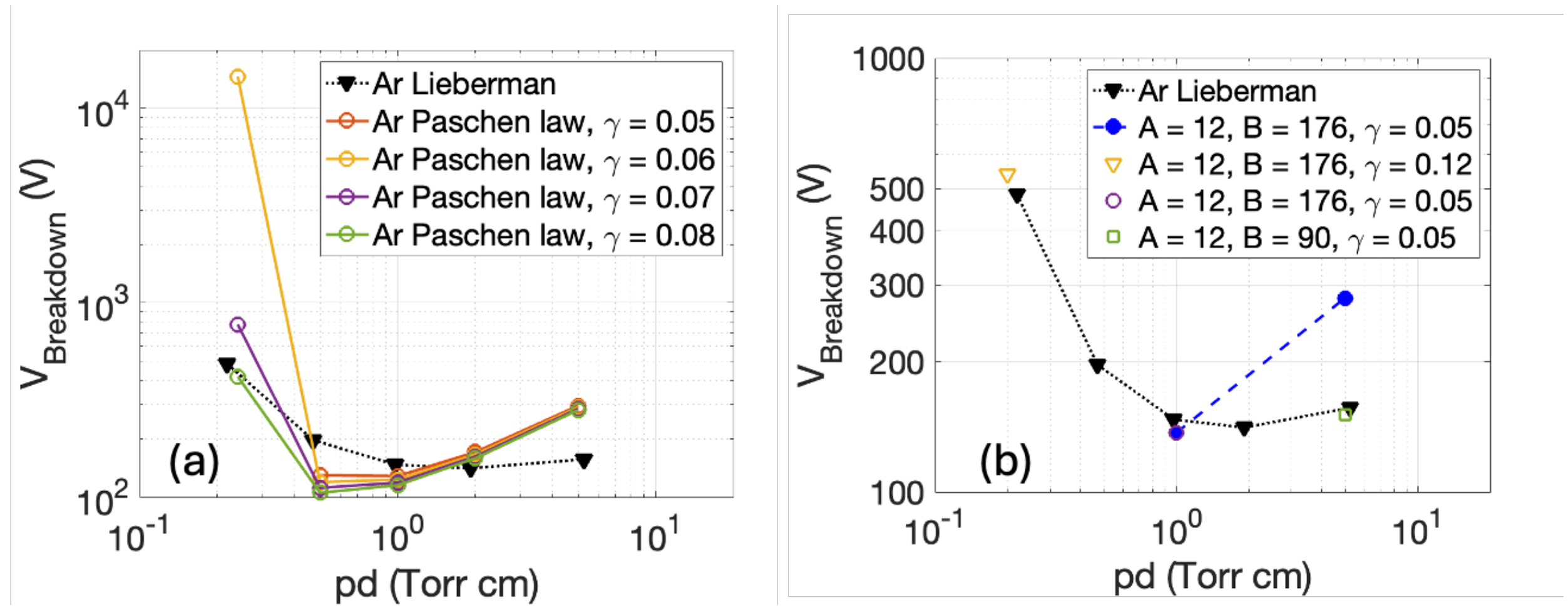

**Figure 7.** Sensitivity of Paschen's law breakdown characteristics to secondary electron emission yield and Townsend ionization parameters for argon at $\mathrm{d} = \mathrm{L} = 1\mathrm{cm}$ and $0.24 < \mathrm{p}\,(\mathrm{Torr}) < 5$. **(a)** Comparison between the experimental Paschen curve reported by [2] and Paschen's law (Eq. (34)) predictions obtained using fixed Townsend coefficients ($\mathrm{A} = 11.5\,\mathrm{cm}^{-1}\,\mathrm{Torr}^{-1}$, $\mathrm{B} = 176\,\mathrm{V\,cm}^{-1}\,\mathrm{Torr}^{-1}$) with varying secondary electron emission yield $\gamma$. The results demonstrate the strong sensitivity of the left-hand branch of the Paschen curve to $\gamma$, particularly at low $\mathrm{pd}$. **(b)** Breakdown voltages obtained using PASCHEN-1D with different combinations of $(\mathrm{A}, \mathrm{B}, \gamma)$, illustrating that increased $\gamma$ is required to reproduce the low-$\mathrm{pd}$ branch, while a reduced effective $\mathrm{B}$ improves agreement on the high-$\mathrm{pd}$ branch. Together, these results highlight the distinct physical roles of secondary emission and gas-phase ionization in shaping the Paschen curve and motivate the use of $\mathrm{pd}$-dependent effective parameters for fluid-based discharge modeling.

The implications of this behavior for fluid-based discharge modeling are illustrated in Fig. 7(b), where the dashed blue curve corresponds to PASCHEN-1D breakdown voltages obtained using a fixed parameter set ($A = 11.5 \text{cm}^{-1}\text{Torr}^{-1}$, $B = 176\ \text{V cm}^{-1}\text{Torr}^{-1}$, and $\gamma = 0.05$). This parameter choice reproduces the breakdown voltage at $pd = 1\ \text{Torr cm}$. However, it fails to produce a breakdown signature on the left branch of the Paschen curve and overestimates the breakdown voltage on the right branch. The absence of the breakdown signature at low $pd$ is notable, as the classical Paschen's law [Eq. (34)] predicts an unphysical negative breakdown voltage of $-147.98\ \text{V}$ at $pd = 0.2\ \text{Torr cm}$ for this parameter set.

Agreement across the full experimental Paschen curve is recovered only when the effective secondary electron emission yield is allowed to vary with $pd$, as indicated by the discrete markers in Fig. 7(b). Specifically, increasing $\gamma$ on the left branch and reducing the effective $B$ value on the right branch yields breakdown voltages that closely match the experimental data. An enhanced $\gamma$ at low $pd$ is consistent with higher ion impact energies at the cathode due to longer ion mean free paths. On the other hand, a reduced effective $B$ at high $pd$ reflects increased ionization efficiency associated with more frequent electron–neutral collisions, resulting in a larger Townsend ionization coefficient $\alpha$.

These results point to a limitation of classical Paschen's law modeling when applied with constant material and plasma discharge parameters and show that reproducing experimental breakdown behavior over a wide range of $pd$ requires accounting for the evolving role of secondary electron emission and ionization processes. PASCHEN-1D provides a flexible framework for exploring these effects self-consistently within a drift–diffusion–Poisson formulation, while remaining compatible with the breakdown theory.

### 4.5 Ultrafast photoemission–driven transient discharge

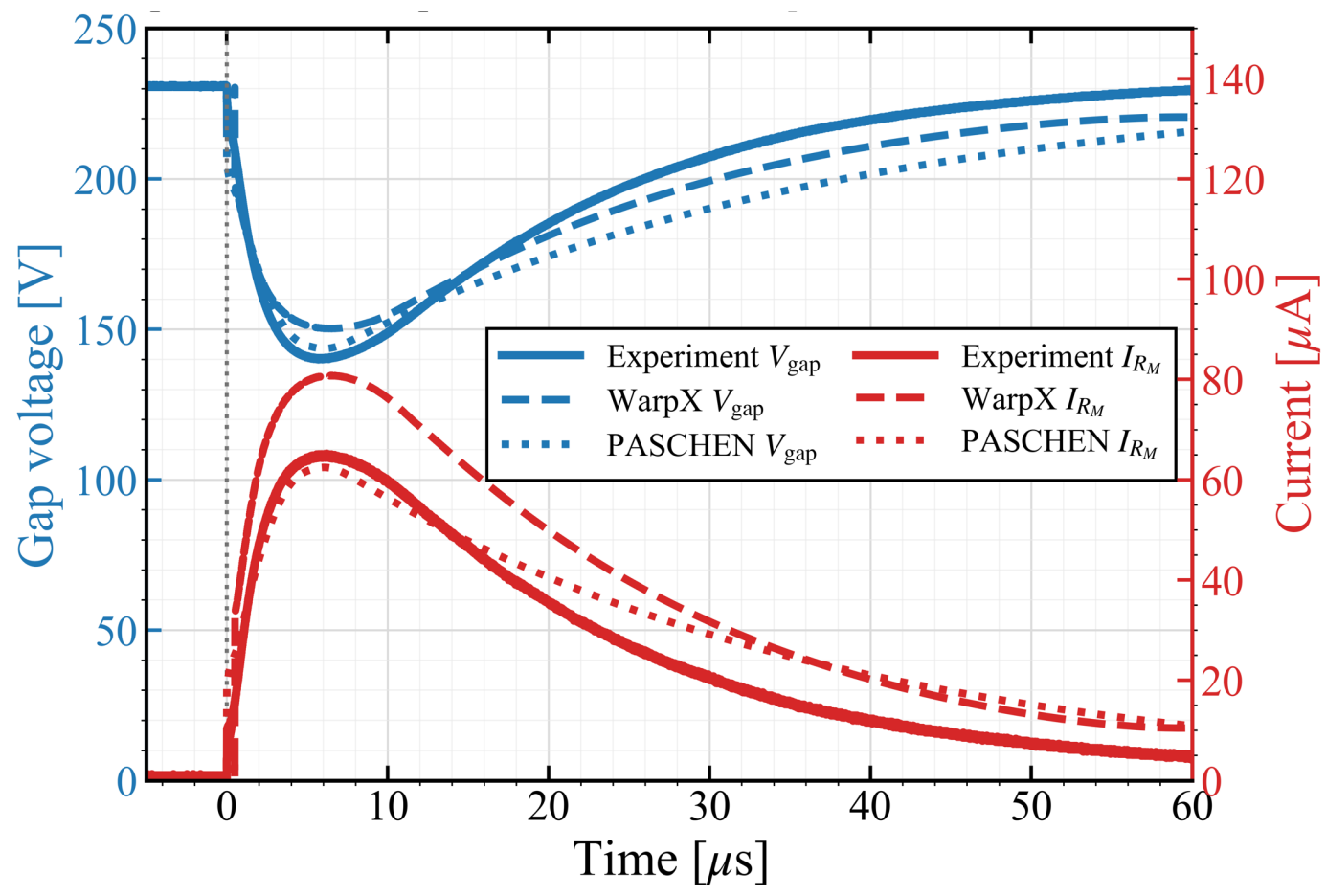

**Figure 8.** Photoemission–driven discharge response obtained from experimental measurements [21] (solid curves), PASCHEN-1D simulation (dotted curves) and 1D WarpX Particle-in-Cell simulation (dashed curve). Blue curves correspond to the plasma gap voltage $V_{gap}$

as functions of time for a dc drive with $V_{app} = 231$ V. Red curves correspond to the discharge current $I_{discharge}(t)$, showing a sharp transient triggered by laser-induced photoemission at $time = 0\mu s$. The cathode photoemission is modeled using a quantum mechanical emission model [44–48] with laser wavelength $\lambda = 230nm$, pulse energy $U = 36\mu J$, pulse duration $\tau_p = 30ps$, and incidence angle $19°$, as reported in [21]. The cathode material Aluminum is characterized by a work function $W = 4.1eV$ and effective Fermi level $E_F = 11.7eV$. The external circuit consists of a source side resistor $R_0 = 400k\Omega$ to model the transient source-voltage sag reported in [21], a series plasma-branch resistor $R_m = 1M\Omega$, and a load side parasitic stray capacitance $C_{ext} = 21pF$. The results demonstrate self-consistent coupling of ultrafast emission, plasma transport, space-charge effects, and circuit dynamics within a unified 1D framework.

Now we consider a discharge driven by an ultrafast time-resolved quantum photoemission from the cathode, representative of ultrafast laser–plasma interaction experiments [23]. In this configuration, the plasma gap is driven by a step voltage of amplitude $V_{app} = 231V$, applied through a resistive–capacitive external network ($R_0 = 400k\ \Omega$, $R_m = 1\ M\Omega$, $C_{ext} = 21\ pF$). An ultrashort laser pulse (pulse duration $\tau_p$=$30ps$) incident on the cathode triggers photoemission, modeled using a quantum mechanical emission model [44–48] that accounts for multiphoton absorption and the electronic density of states of the emitter. The emission current density $J_{emit}(t)$ is precomputed on a refined temporal grid and injected self-consistently as a cathode boundary flux.

Figure 8 shows the resulting voltage and current dynamics. Prior to laser excitation, the discharge relaxes toward a quasi-steady state determined by the applied voltage and external circuit. At the time of laser arrival ($time\ =\ 0\mu s$ in Fig. 8), the rapid injection of photoemitted electrons produces a sharp transient increase in discharge current, accompanied by a rapid drop in the plasma gap voltage due to enhanced space-charge screening. Following the pulse, the system recovers on a slower time scale toward the pre-pulse operating point, governed by plasma transport and circuit recharging. These trends recovered by PASCHEN-1D simulation (dotted curves in Fig. 8) are in agreement with time-resolved ultrafast photoemission-driven discharge experiments (solid curves in Fig. 8) reported in Ref. [21] as well as independent Particle-in-Cell simulations (dashed curves in Fig. 8) using WarpX [55,56].

This example demonstrates several capabilities of PASCHEN-1D. The solver robustly bridges picosecond-scale emission physics and microsecond-scale plasma–circuit dynamics within a single time-domain framework. It also enables direct coupling between cathode physics, sheath formation, and global discharge behavior. The same infrastructure accommodates alternative emission models, as described in Section 2.8, without modification of the core plasma or circuit solvers.

## 5 Scope and avenues for future improvement

To place PASCHEN-1D in context, Table 1 summarizes its modeling scope alongside several widely used fluid and kinetic plasma tools. The purpose of this comparison is not to assess overall solver generality or fidelity, but to clarify the class of problems for which PASCHEN-1D is intended. The framework is designed for one-dimensional breakdown and discharge-transition studies in which cathode emission, dielectric charging, and external circuit coupling play a central role, features that are often simplified or treated separately in other codes.

Table 1. Contextual comparison of PASCHEN-1D with representative open-source plasma modeling tools

| Code | Approach & dimensionality | Primary physics scope | External circuit / emission coupling | Open-source |
|---|---|---|---|---|
| PASCHEN-1D (this work) | Drift–diffusion–Poisson fluid; 1D. Central (Kurganov-Tadmor) non-oscillatory flux with explicit diffusion | Gas breakdown, Townsend–glow transition, pulsed, DC, and RF discharges; dielectric charging; ultrafast plasma phenomena | Self-consistent electrode emission (secondary, photo, field, thermionic) and unified external R/L/C network | Yes |
| Zapdos (+MOOSE) [23] | Finite-element fluid; 1D/2D/3D | General low-temperature plasma transport and chemistry | Limited / user-implemented | Yes |
| Afivo-streamer [26] | Fluid with Adaptive Mesh Refinement (AMR); 1D/2D/3D | Streamer and fast ionization wave dynamics [57] (DBD, leaders) | No explicit circuit or emission framework | Yes |
| SOMAFOAM (OpenFOAM-based) [24] | Finite-volume fluid; 1D/2D/3D | Continuum LTP modeling including plasma–dielectric | External circuit typically | Yes |
| | | interaction; DC/RF/microwave | treated indirectly | |
| EDIPIC [27] | PIC/MCC; 1D/2D | Kinetic low-temperature plasmas (electrostatic, RF) | Limited external circuit models; kinetic emission | Yes |
| XOOPIC / OOPIC [28] | PIC (ES/EM); 2D | General kinetic plasmas | Basic circuit coupling; kinetic emission | Yes |
| ZDPlasKin (0D) [25] | Global (volume-averaged) chemistry | Time-resolved species kinetics and gas temperature | Not applicable | No (freeware) |
| WarpX , Warp [55,56] | PIC / PIC–AMR; 1D/2D/3D (ES/EM) | Kinetic plasma dynamics, beam–plasma interaction, space-charge–dominated discharges, RF/microwave, and accelerator plasmas | Limited / user-implemented (basic external circuits; kinetic emission models) | Yes |
| BOLSIG+ [30] | Boltzmann solver (EEDF); 0D | Swarm parameters and transport coefficients | Not applicable | No (binary only) |

Despite the broad range of benchmark cases presented in this work, several limitations of the current PASCHEN-1D framework should be noted. These limitations arise both from the underlying one-dimensional fluid formulation and from practical modeling choices that affect quantitative agreement with experiments.

A persistent challenge in fluid-based discharge modeling is the selection of transport (mobility and diffusion), and rate (ionization and recombination) coefficients. Reported values for mobilities, diffusion coefficients, and Townsend parameters are typically based on empirical fits and vary across textbooks, databases, and publications, even for the same gas species. In addition, available data often provide incomplete coverage across relevant E/p regimes, which can strongly influence breakdown thresholds, transient growth rates, and steady-state plasma properties. Even when coefficients are derived from swarm measurements or cross-section databases [30–32], their direct use may not necessarily yield quantitative agreement with a given experimental configuration, due to differences in geometry, surface conditions, operating regime, and the often less-considered circuit configurations. As seen in the Paschen curve benchmarks, reproducing breakdown behavior over a wide range of $pd$ requires careful parameter selection, and no single set of coefficients is guaranteed to reproduce all regimes simultaneously. This sensitivity reflects a broader limitation of fluid models that rely on tabulated or fitted coefficients rather than self-consistent kinetic evolution.

More fundamentally, PASCHEN-1D shares the intrinsic limitations of drift–diffusion fluid models when compared with kinetic approaches such as particle-in-cell (PIC) simulations. The model assumes locally equilibrated transport and does not explicitly resolve electron or ion velocity distribution functions. As a result, non-Maxwellian effects [2], high-energy electron tails, and strongly nonlocal transport [1,22] are represented only through effective transport and ionization coefficients. Although extended fluid formulations, e.g., energy-equation closures [58,59], nonlocal transport models [60], or moment-based Boltzmann solvers [61], have been proposed in the literature, these approaches are not yet implemented in PASCHEN-1D and remain a natural direction for future development.

The present implementation treats the plasma as a two-fluid system consisting of electrons and a single effective ion species. While this assumption is sufficient for the benchmark cases considered here, many practical discharges require a more detailed description, including multiple ion species, metastables, and extended reaction pathways. Incorporating additional species and reaction networks would enable improved modeling of gas-specific chemistry, stepwise ionization, and energy transfer processes, particularly in molecular gases and complex mixtures.

An additional area of ongoing effort concerns capacitively coupled plasma (CCP) discharges. PASCHEN-1D includes the core ingredients required for CCP modeling, i.e., self-consistent plasma–circuit coupling, dielectric effects, and flexible waveform excitation. However, CCP discharges are especially sensitive to nonlocal electron kinetics, sheath dynamics, and energy-dependent transport coefficients, which are known to be difficult to capture accurately within drift–diffusion fluid models. A systematic CCP validation effort, including direct comparison with PIC/MCC simulations and published benchmarks, is therefore deferred to future work.

These limitations define the current scope of PASCHEN-1D rather than detract from the benchmark results presented in this work. The solver is intentionally modular, allowing future extensions such as improved transport closures, energy-based formulations, and multi-species chemistry to be incorporated while preserving the validated plasma–surface–circuit coupling demonstrated here.

## 6. Conclusions

We have presented PASCHEN-1D, a one-dimensional drift–diffusion–Poisson plasma simulation framework. The model integrates self-consistent electron and ion transport with dielectric charging, heterogeneous electrode emission mechanisms (e.g., ion-induced secondary emission, (ultrafast) photoemission, field emission, and thermionic emission), and flexible external circuit coupling.

Numerically, PASCHEN-1D employs a Kurganov–Tadmor central upwind scheme with slope limiting for drift transport, explicit diffusion handling, fourth-order Runge–Kutta time integration, and a fast tridiagonal Poisson solver coupled to implicit boundary iterations. A dedicated convergence study confirms numerical consistency under systematic spatial and temporal refinement, ensuring that the reported results reflect physical behavior rather than discretization artifacts.

The model has been validated against multiple benchmark problems, including nanosecond pulsed dielectric-barrier discharges, DC Townsend-to-glow transitions, breakdown identification via negative differential resistance, Paschen curve reconstruction for argon and nitrogen, and ultrafast photoemission induced plasmas. Across these cases, PASCHEN-1D reproduces key experimental trends, including breakdown thresholds, transient plasma formation, dielectric screening, and regime transitions over picosecond-to-microsecond time scales. The Paschen curve analysis illustrates the strong sensitivity of breakdown behavior over a wide range of $pd$ to the evolving role of secondary electron emission and ionization processes.

The modular architecture of PASCHEN-1D provides a robust foundation for future extensions such as energy-resolved or nonlocal transport closures, multi-species reaction networks, and hybrid kinetic–fluid coupling for regimes where velocity distribution effects become dominant. Planned developments also include the incorporation of magnetic fields and extension to higher-dimensional geometries, enabling the treatment of transverse dynamics and spatial nonuniformities. Validation against capacitively coupled plasma configurations is also planned. Within its current scope, PASCHEN-1D provides a computationally stable and efficient framework for modeling gas breakdown, discharge initiation, and plasma–circuit interaction in DC and pulsed systems.

This work was supported by National Science Foundation (NSF) Grant No. 2516752, the Air Force Office of Scientific Research (Grants No. FA9550-20-1-0409 and No. FA9550-22-1-0523), and Department of Energy (DOE) Basic Energy Sciences (BES) through

Award No. DE-SC0026271.

During the preparation of this work the author(s) used ChatGPT (developed by OpenAI) in order to assist with grammatical editing and refinement of scientific language. After using this tool/service, the author(s) reviewed and edited the content as needed and take(s) full responsibility for the content of the published article.

# Appendix A: Verification of External Circuit Coupling and Component Behavior

This appendix provides a systematic verification of the external circuit coupling implemented in PASCHEN-1D by isolating and independently sweeping individual lumped circuit elements while holding all plasma and geometric parameters fixed. The primary objective is not to introduce new physical results, but to demonstrate that each circuit component, e.g., series and parallel resistances, capacitances, and inductances, produces the expected response when coupled self-consistently to the plasma discharge. The presented tests span increasingly complex circuit topologies, from simple $R_0$-$C_p$ networks to $R_0$-$C_s$-$L_s$-$C_p$-$L_p$-$R_m$ configurations and examine their impact on the plasma gap voltage and discharge current. The observed trends are consistent with classical circuit theory, confirming the correctness, stability, and generality of the unified plasma–circuit formulation employed in PASCHEN-1D.

### A.1 Effect of shunt capacitance $C_p$ in an $R_0$-$C_p$ network

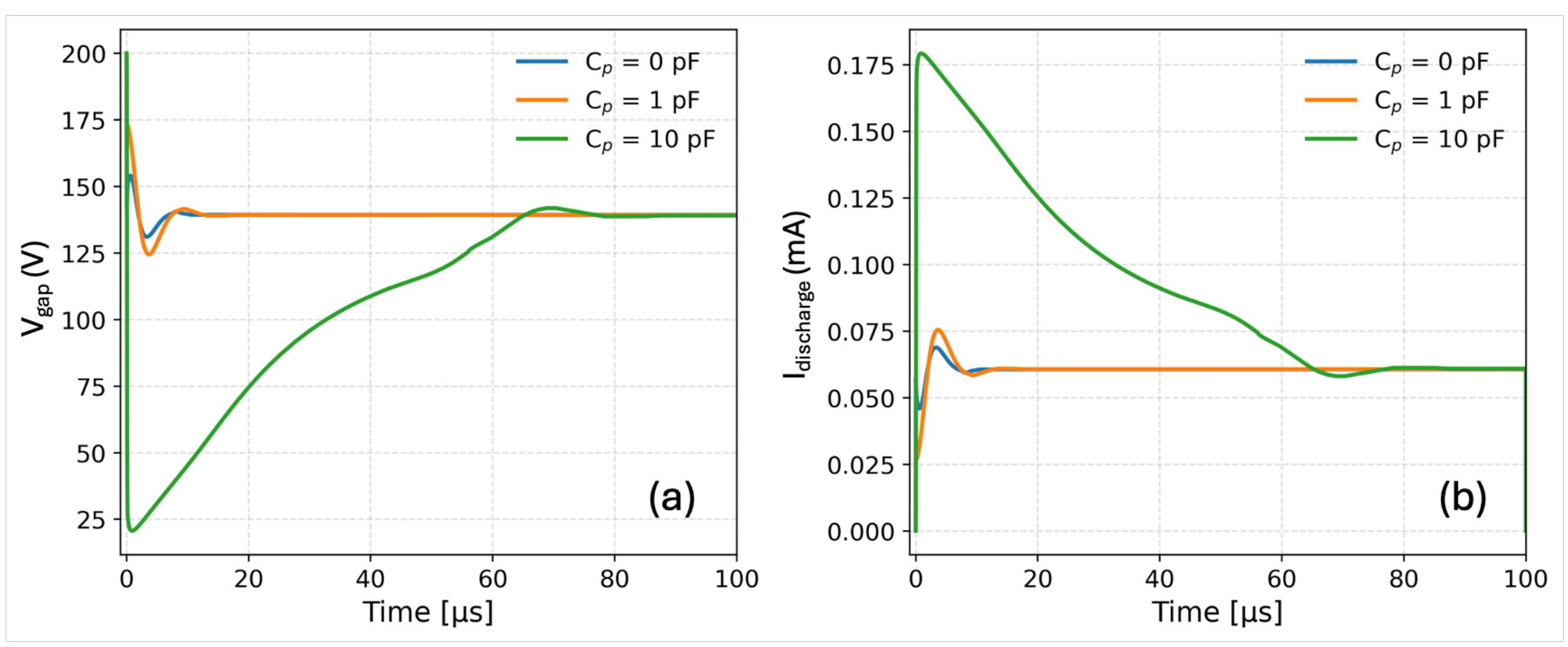

**Figure A.1.** Effect of shunt capacitance $C_p$ on the transient plasma–circuit response in an "R0_Cp" configuration. The applied voltage is a 200 V step, with $R_0$=1MΩ, argon at p=1 Torr, and L=1 cm. (a) Time evolution of the plasma gap voltage $V_{gap}(t)$ for $C_p$=0, 1, and 10 pF. Increasing $C_p$ produces a stronger initial voltage sag and delayed recovery due to enhanced displacement current at the plasma-side node. (b) Corresponding discharge current $I_{discharge}(t)$, showing larger early-time current transients for larger $C_p$, followed by convergence to a common steady-state current. The results confirm that the shunt capacitance primarily affects transient dynamics, while the long-time operating point is governed by the plasma conductivity and series resistance.

### A.2 Effect of plasma-branch series resistance $\mathrm{R_m}$ in an $\mathrm{R_0}$-$\mathrm{C_p}$-$\mathrm{R_m}$ network

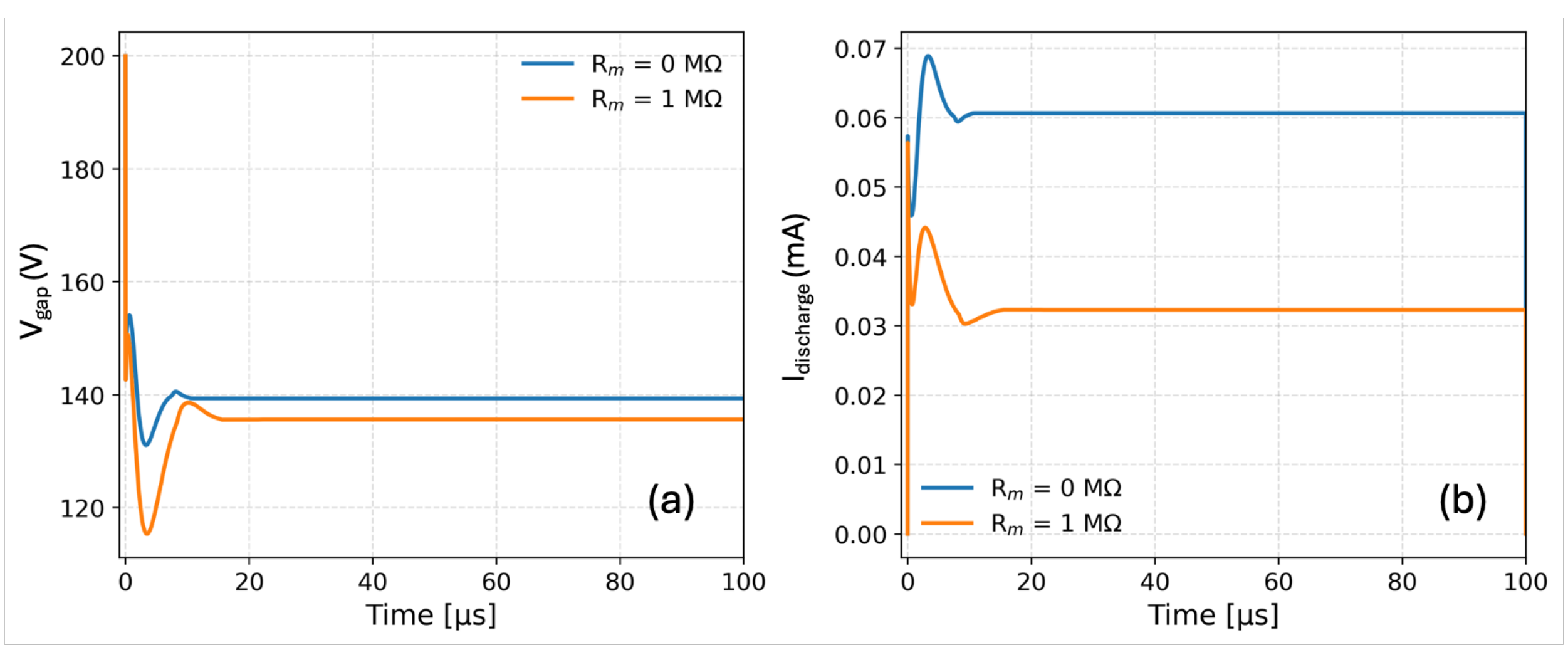

**Figure A.2.** Effect of plasma-branch series resistance $\mathrm{R_m}$ in an "R0_Cp_Rm" configuration. The discharge is driven by a 200 V step through a series resistor $\mathrm{R_0}$=1MΩ with a shunt capacitance $\mathrm{C_p}$=1pF; all remaining plasma and geometry parameters are identical to Fig. A.1. Results are shown for $\mathrm{R_m}$= 0 and 1MΩ. (a) Gap-voltage response $\mathrm{V_{gap}(t)}$. (b) Discharge current $\mathrm{I_{discharge}(t)}$. Introducing $\mathrm{R_m}$ increases the effective series impedance of the plasma branch, reducing the steady-state discharge current and producing a lower quasi-steady operating point while preserving the early-time capacitive transient. The trends confirm that PASCHEN-1D captures the expected redistribution between displacement-current–dominated startup dynamics and the resistively limited long-time conduction state when explicit plasma-branch impedance is added.

### A.3 Effect of series coupling capacitor $\mathrm{C_s}$ in an $\mathrm{R_0}$-$\mathrm{C_s}$-$\mathrm{C_p}$ network

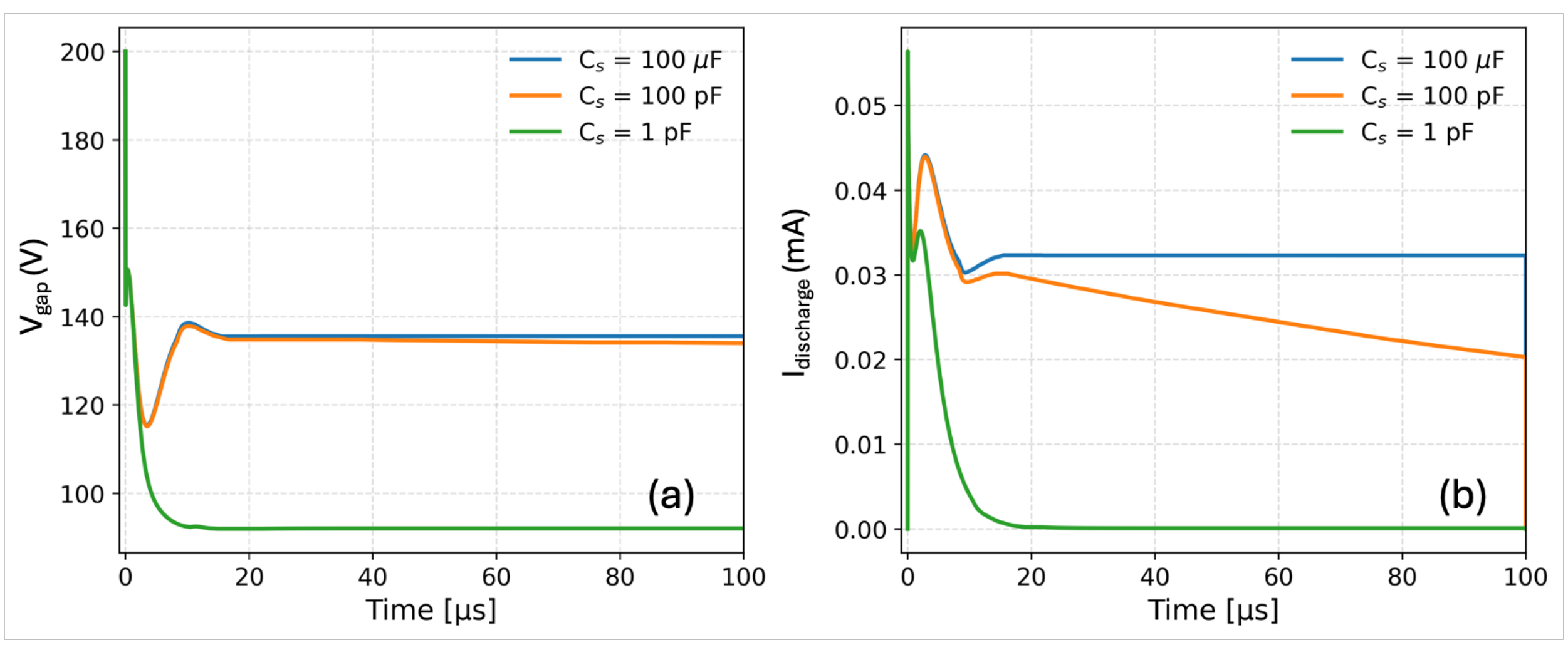

**Figure A.3.** Effect of the series coupling capacitor $C_s$ in an "R0_Cs_Cp_Rm" configuration. The discharge is driven by a 200 V step through a series resistor $R_0$=1 MΩ, with a shunt capacitance $C_p$=1pF at the plasma-side node and no additional plasma-branch resistance ($R_m$=0). Results are shown for $C_s$=100μF, 100pF, and 1pF; all plasma, geometry, and gas parameters are identical to Figs. A.1–A.2. (a) Plasma gap voltage $V_{gap}(t)$. (b) Discharge current $I_{discharge}(t)$. Decreasing $C_s$ increasingly suppresses long-time current conduction, transitioning the system from quasi-DC coupling ($C_s \rightarrow \infty$) to strongly capacitive coupling, where only a transient displacement-current response is sustained. The observed trends confirm that PASCHEN-1D correctly captures series capacitive isolation and its impact on plasma–circuit energy transfer.

## A.4 Effect of series inductance $L_s$ in an $R_0$-$C_s$-$L_s$-$C_p$-$R_m$ network

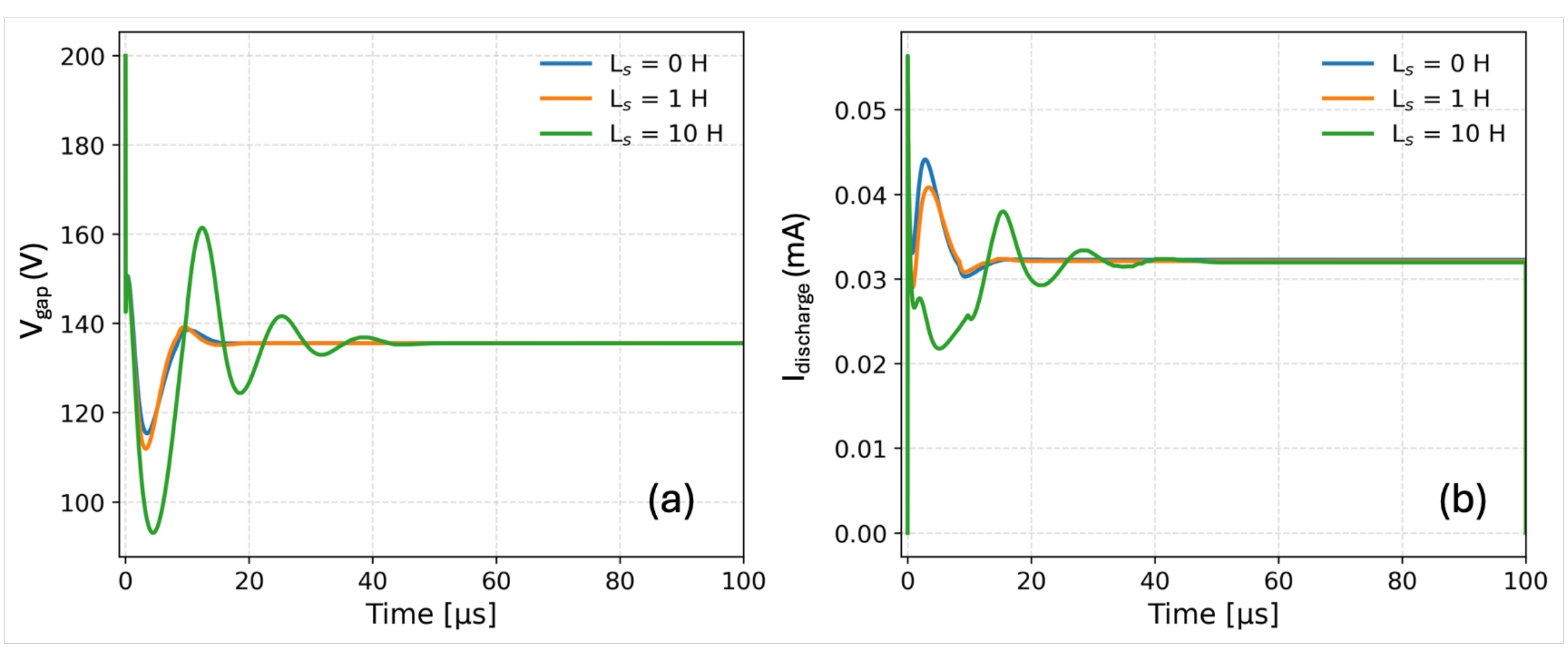

**Figure A.4.** Effect of the series inductance $L_s$ in an "R0_Cs_Ls_Cp_Rm" configuration. The discharge is driven by a 200 V step through a source resistance $R_0$=1MΩ, a series capacitor $C_s$=100μF, and a plasma-side shunt capacitor $C_p$=1pF, with a plasma-branch resistance $R_m$=1MΩ. Results are shown for $L_s$=0, 1, and 10 H; all plasma, geometry, and gas parameters are identical to Figs. A.1–A.3. (a) Plasma gap voltage $V_{gap}(t)$. (b) Discharge current $I_{discharge}(t)$. Increasing $L_s$ introduces pronounced transient oscillations and delays the establishment of the steady-state operating point, reflecting the expected inductive impedance and energy storage effects. The damping of these oscillations by the resistive and plasma elements confirms correct inductive coupling and stable integration of the circuit–plasma system in PASCHEN-1D.

## A.5 Effect of parallel inductance $L_p$ in an $R_0$-$C_s$-$L_s$-$C_p$-$L_p$-$R_m$ network

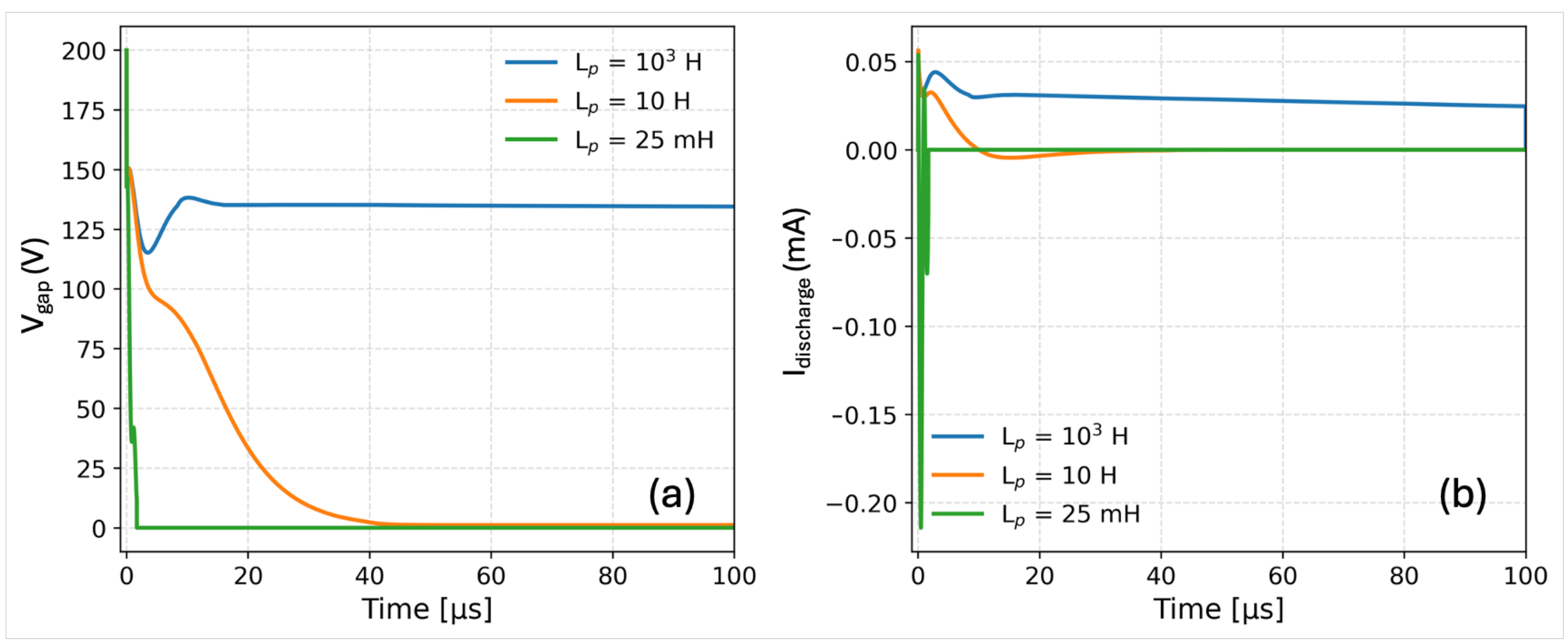

**Figure A.5.** Effect of the parallel inductance $L_p$ in an "`R0_Cs_Ls_Cp_Lp_Rm_Cext`" configuration. The discharge is driven by a 200 V step through a source resistance $R_0$=1MΩ, a series capacitor $C_s$=100 µF, a series inductor $L_s$=0.01H, and a plasma-side shunt capacitor $C_p$=1pF, with a plasma-branch resistance $R_m$=1MΩ. Results are shown for $L_p$=25mH, 10 H, and $10^3$ H; all plasma, geometry, and gas parameters are identical to Figs. A.1–A.4. (a) Plasma gap voltage $V_{gap}(t)$. (b) Discharge current $I_{discharge}(t)$. Decreasing $L_p$ progressively diverts current away from the plasma branch, suppressing the gap voltage and driving the system toward a short-circuit response, while large $L_p$ effectively isolates the plasma branch and recovers the behavior of a purely capacitive shunt. The observed trends confirm correct implementation of parallel inductive coupling in PASCHEN-1D.